\documentclass[1p]{elsarticle}

\usepackage{amssymb}
\usepackage{amsfonts}
\usepackage{amstext}
\usepackage{graphicx}
\usepackage{amscd}
\usepackage{amsmath}
\usepackage{mathrsfs}
\usepackage{stmaryrd}
\usepackage{enumerate}

\newcommand{\ihbar}{\imath \hbar}
\newcommand{\Obj}{\mathrm{Obj}}
\newcommand{\Morph}{\mathrm{Morph}}
\newcommand{\id}{\mathrm{id}}
\newcommand{\Aut}{\mathrm{Aut}}
\newcommand{\Funct}{\mathrm{Funct}}
\newcommand{\Ad}{\mathrm{Ad}}
\newcommand{\soplus}{\inplus}
\newcommand{\dd}{d_{(2)}}
\newcommand{\PP}{\mathbb{P}}

\newcommand{\Skel}{\mathrm{Skel}}
\newcommand{\dist}{\mathrm{dist}_{\mathrm{FS}}}
\newcommand{\Ran}{\mathrm{Ran}\, }
\newcommand{\Te}{\mathbb{T}e}
\newcommand{\tr}{\mathrm{tr}}
\newcommand{\Span}{\mathrm{Span}}

\newtheorem{definition}{Definition}
\newtheorem{property}{Property}
\newtheorem{theorem}{Theorem}
\newenvironment{proof}{Proof:}{$\Box$}
\newenvironment{example}{Example:}{}
\newtheorem{proposition}{Proposition}
\newtheorem{lemma}{Lemma}

\journal{Journal of Geometry and Physics}

\begin{document}
\begin{frontmatter}

\title{Non-abelian higher gauge theory and categorical bundle}  

\author[uti]{David Viennot}
\address[uti]{Institut UTINAM (CNRS UMR 6213, Universit\'e de Franche-Comt\'e, Observatoire de Besan\c on), 41 bis Avenue de l'Observatoire, 25000 Besan\c con cedex, France.}

\begin{abstract}
A gauge theory is associated with a principal bundle endowed with a connection permitting to define horizontal lifts of paths. The horizontal lifts of surfaces cannot be defined into a principal bundle structure. An higher gauge theory is an attempt to generalize the bundle structure in order to describe horizontal lifts of surfaces. A such attempt is particularly difficult for the non-abelian case. Some structures have been proposed to realize this goal (twisted bundle, gerbes with connection, bundle gerbe, 2-bundle). Each of them uses a category in place of the total space manifold of the usual principal bundle structure. Some of them replace also the structure group by a category (more precisely a Lie crossed module viewed as a category). But the base space remains still a simple manifold (possibly viewed as a trivial category with only identity arrows). We propose a new principal categorical bundle structure, with a Lie crossed module as structure groupoid, but with a base space belonging to a bigger class of categories (which includes non-trivial categories), that we called affine 2-spaces. We study the geometric structure of the categorical bundles built on these categories (which are a more complicated structure than the 2-bundles) and the connective structures on these bundles. Finally we treat an example interesting for quantum dynamics which is associated with the Bloch wave operator theory.
\end{abstract}

\end{frontmatter}

\tableofcontents

\section{Introduction}
The geometry of the principal bundles plays an important role in theoretical physics. It is the natural framework to model the fundamental interactions between point particles in classical field theory, and is the startpoint for the quantum field theory \cite{naber}. Moreover in nonrelativistic quantum physics, the geometric (Berry) phase phenomenon \cite{berry} is closely related to this geometry. A principal bundle naturally arises to treat cyclic quantum dynamics \cite{aharonov} or adiabatic quantum dynamics driven by classical parameters \cite{simon,wilczek,sardanashvily,viennot1}. These physical problems are associated with the holonomies or the horizontal lifts of paths drawn on the base manifold of the principal bundle.\\
The horizontal lifts of surfaces cannot be defined within the framework of the principal bundles. The interest for the horizontal lifts of surfaces arises from the development of the string and brane theories, in which the string and brane gauge theory is associated with holonomies of surfaces \cite{kapustin,kalkkinen,murray,aschieri,gawedzki}. Recently, we have shown that the geometric phases associated with quantum systems submitted to some decoherence processes take place in higher gauge theories associated with horizontal lifts of surfaces \cite{viennot2,viennot3,viennot4}.\\
Geometric realizations of the abelian higher gauge theories are well understood, as gerbes with connection \cite{brylinski}, bundle gerbes \cite{murray2} or twisted bundles \cite{mackaay}. For the non-abelian higher gauge theories, some generalizations of these geometric realizations have been proposed: non-abelian gerbes with connection \cite{laurent,breen}, non-abelian bundle gerbes \cite{kalkkinen2, aschieri2}, non-abelian twisted bundles \cite{aschieri2} and parallel transport over path spaces \cite{cattaneo,chatterjee,chatterjee2}. We can also cite the higher gauge structure arising in the study of the principal composite bundles \cite{viennot5}. In these approaches, the total space of the geometric structure is not a smooth manifold as in the usual gauge theory but a category. For some of these approaches, the structure group is also replaced by a structure which can be viewed as a category (as for example an extension of Lie groups). Another interesting approach of non-abelian higher gauge theories has been proposed by Baez \textit{et al} and Wockel, the 2-bundles \cite{baez1,baez2,baez3,wockel,soncini}. In this approach, the structure group is replaced by a Lie crossed module. The different approaches seem to be equivalent \cite{nikolaus}. The strategy followed by Baez \textit{et al} to define the 2-bundles is very interesting since it is based on the idea consisting to substitute at each smooth manifold a geometric category called a 2-space. Unfortunately this goal seems unachieved since in the 2-bundle theory the base space is restricted to the trivial 2-spaces (i.e. an usual manifold $M$ considered as a category $\mathcal M$ with $\Obj(\mathcal M) = M$ and $\Morph(\mathcal M) = \{\id_x\}_{x \in M}$). The reason of this restriction is the difficulty to define a ``2-cover'' of a 2-space. Indeed the definition of the union of two open sub-2-spaces is not clear since it needs to know how to compose an arrow in one 2-space with an arrow in another one. The trivial 2-spaces are very poor categories since they have only identity arrows.\\
In this paper, inspirated by the 2-bundle theory, we propose a new theory of categorical bundles with a bigger class of 2-spaces that we call the affine 2-spaces. It includes the trivial 2-spaces but also categories with non-trivial arrows. Moreover the union of two affine open sub-2-spaces is clearly defined. The affine 2-spaces are introduced in the next section. Section 3 introduces the 2-bundles over affine 2-spaces and explores their algebraic and geometric properties. In particular we show that the structure is very more rich than the usual 2-bundle theory, since a new kind of 1-transition functions appears. 2-bundles over affine 2-spaces are endowed with connective structures in section 4, and the horizontal lifts are considered in section 5. Finally section 6 presents a simple physical example based on the use of the Bloch wave operators and their generalizations in quantum dynamics.\\

\textit{A note about the notations used here: let $A$ be a category, $\Obj(A)$ denotes its set of objects, $\Morph(A)$ denotes its set of arrows (so called morphisms), $s: \Morph(A) \to \Obj(A)$ denotes its source map, $t:\Morph(A)\to \Obj(A)$ denotes its target map, $\circ : \Morph(A) \times_{s=t} \Morph(A) \to \Morph(A)$ denotes the composition of the arrows and $\id : \Obj(A) \to \Morph(A)$ denotes its identity map. Let $G$ be a Lie group, $e_G$ denotes its neutral element, $\Aut(G)$ denotes its group of automorphisms and $\mathrm{Der}(\mathfrak g)$ denotes the algebra of the derivations of its Lie algebra $\mathfrak g$. Let $M$ be a differential manifold, $\underline G_M$ denotes the set of $\mathcal C^\infty$ functions from $M$ to $G$, $TM$ denotes the tangent space of $M$ and $\Omega^n(M,X)$ denotes the set of $X$ valued differential $n$-forms of $M$. Let $P$ be a principal bundle over $M$, $HP$ denotes the horizontal tangent space of $P$, $VP$ denotes the vertical tangent space of $P$ and $\Gamma(M,P)$ denotes the set of the sections from $M$ to $P$.}

\section{Affine 2-spaces}
\begin{definition}[2-space]
A smooth 2-space is a category $\mathcal M$ such that $\Obj(\mathcal M)$ and $\Morph(\mathcal M)$ are smooth manifolds, and such that $s,t:\Morph(\mathcal M) \to \Obj(\mathcal M)$, $\id:\Obj(\mathcal M) \to \Morph(\mathcal M)$ and $\circ : \Morph(\mathcal M)_s\times_t \Morph(\mathcal M) \to \Morph(\mathcal M)$ are smooth maps.
\end{definition}

\begin{definition}[Affine space]
An affine space is defined by three kinds of data $(M,E,\varphi)$ where $M$ is a manifold, $E$ is a vector space and $\varphi : M^2 \to E$ is an application such that
\begin{enumerate}[i.]
\item $\forall x \in M$, $\varphi(x,x) = \overrightarrow 0$.
\item $\forall x,y,z \in M$, $\varphi(x,y)+\varphi(y,z) = \varphi(x,z)$.
\item $\forall x \in M$,$\forall \overrightarrow u \in E$, $\exists ! y \in M$ such that $\varphi(x,y) = \overrightarrow u$
\end{enumerate}
\end{definition}
An affine space is generally the consideration of a flat manifold where we indentify each of their tangent spaces with the set of bipoints. We want to extend this notion to more general situations.

\begin{definition}[Affine 2-space]
We call affine 2-space the three kinds of data $(\mathcal M,\mathcal R,\varphi)$ where the category $\mathcal M$ is a 2-space, $\mathcal R$ is a reflexive and symetric relation on $\Obj(\mathcal M)$ (if $x \mathcal R y$ we say that $x$ and $y$ are linkable) and $\varphi : \bigsqcup_{n \in \mathbb N^*} \Obj(\mathcal M)^n_{/\mathcal R} \to \Morph(\mathcal M)$ is a surjective map where
$$ \Obj(\mathcal M)^n_{/\mathcal R} = \{(x_n,...,x_1) \in \Obj(\mathcal M)^n | \forall i<n, x_{i+1} \mathcal R x_i \} $$
An affine 2-space is such that
\begin{enumerate}[i.]
\item[0.] $\forall (y,...,x)\in \Obj(\mathcal M)^n_{/\mathcal R}$, $s(\varphi(y,...,x))=x$ and $t(\varphi(y,...,x))=y$.
\item $\forall x \in \Obj(\mathcal M)$, $\varphi(x) = \id_x$ and $\forall (...,x,x,...) \in \Obj(\mathcal M)^n_{/\mathcal R}$, $\varphi(...,x,x,...)=\varphi(...,x,...)$.
\item $\forall (y,...,x) \in \Obj(\mathcal M)^n_{/\mathcal R}$, $\forall (z,...,y) \in \Obj(\mathcal M)^p_{/\mathcal R}$, $\varphi(z,...,y)\circ\varphi(y,...,x)=\varphi(z,...,y,...,x)$.
\item $\forall x \in \Obj(\mathcal M)$, $\forall f \in \Morph(\mathcal M)$ with $s(f)=x$, $\exists n \in \mathbb N^*$, $\exists (z,...,x) \in \Obj(\mathcal M)^n_{/\mathcal R}$ such that $\varphi(z,...,x) = f$. Moreover, if $n_0 = \min\{n \in \mathbb N^* | \exists  (z,...,x) \in \Obj(\mathcal M)^n_{/\mathcal R} \text{ such that } \varphi(z,...,x) = f \}$ then there exists only one $(z,...,x) \in \Obj(\mathcal M)^{n_0}_{/\mathcal R}$ such that $\varphi(z,...,x)=f$.
\end{enumerate}
\end{definition}
The assumptions i., ii. and iii. are weaker versions of the corresponding assumptions in the definition of an affine space. If $\forall x,y \in \Obj(\mathcal M)$ we have $x\mathcal R y$ we say that $\mathcal M$ is totally linkable.\\
There are three important kinds of affine 2-space.
\begin{definition}[Euclidean affine 2-space]
An affine 2-space is said euclidean if $\mathcal R$ is transitive ($\mathcal R$ is then an equivalence relation) and if
$$ \forall (y,...x) \in \Obj(\mathcal M)^n_{/\mathcal R}, \quad \varphi(y,...,x) = \varphi(y,x) $$
\end{definition}
In the euclidean affine 2-space, the Chasles relation takes the same form than in the affine spaces
\begin{equation}
\forall (x,y,z) \in \Obj(\mathcal M)^3_{/\mathcal R}, \quad \varphi(z,y)\circ\varphi(y,x) = \varphi(z,x)
\end{equation}
Moreover there is a bijection between $\Morph(\mathcal M)$ and $\Obj(\mathcal M)^2_{/ \mathcal R}$.

\begin{definition}[Spherical affine 2-space]
An affine 2-space is said spherical if $x\mathcal R y \Rightarrow x=y$.
\end{definition}
In a spherical affine 2-space the set of the arrows is reduced to $\Morph(\mathcal M) = \{\id_x, x \in \Obj(\mathcal M) \}$ (a spherical affine 2-space is generally called a trivial 2-space).

\begin{definition}[Hyperbolic affine 2-space]
An affine 2-space is said hyperbolic if $\forall n \in \mathbb N^*$, $\Obj(\mathcal M)^n_{/\mathcal R} \not= \varnothing$ and $\forall (y_1,...,x_1),(y_2,...,x_2) \in \bigsqcup_n \Obj(\mathcal M)^n_{/\mathcal R}$, $\varphi(y_1,...,x_1) = \varphi(y_2,...,x_2) \Rightarrow (y_1,...,x_1) \sim (y_2,...,x_2)$ where $\sim$ signifies that the two sequences are equal modulo consecutive repetitions ($(...,x,x,...) \sim (...,x,...)$). 
\end{definition}
In a hyperbolic affine 2-space, there is a bijection between the set of the arrows and the set of the sorted collections of linkable objects without consecutive repetitions. We restrict our attention on these three cases.\\
We note that an affine space $(M,E,\varphi)$ can be viewed as a totally linkable euclidean affine 2-space $\mathcal M$ with $\Obj(\mathcal M) = M$ and $\Morph(\mathcal M) = M \times E$ with $s(x,\overrightarrow u) = x$, $t(x,\overrightarrow u) = y$ such that $\varphi(x,y)=\overrightarrow u$, $\id_x = (x,\overrightarrow 0)$ and $(\varphi(x,\overrightarrow u),\overrightarrow v) \circ (x,\overrightarrow u) = (x, \overrightarrow u+\overrightarrow v)$.\\
The justification of the adjectives euclidean, spherical and hyperbolic is the following. Let $\Obj(\mathcal M)$ be $\mathbb R^2$, the sphere $S^2$ or the Poincar\'e hyperbolic plane. Let $(D)$ be a geodesic of $\Obj(\mathcal M)$ and $\mathcal R$ be such that $x\mathcal R y$ if and only if the geodesic joining $x$ and $y$ is parallel and not confused to $(D)$. Let $\Morph(\mathcal M)$ be the set of the oriented piecewise geodesic paths with edges parallel and not confused to $(D)$. The affine 2-space $\mathcal M$ is then:
\begin{itemize}
\item euclidean if $\Obj(\mathcal M)$ is the plane (since it exists only one geodesic parallel to $(D)$ and passing through a point $x \not\in (D)$); 
\item spherical if $\Obj(\mathcal M)$ is the sphere (since it does not exit a geodesic parallel to $(D)$ and passing through a point $x \not\in (D)$); 
\item hyperbolic if $\Obj(\mathcal M)$ is the Poincar\'e plane (since it exists an infinity of geodesics parallel to $(D)$ and passing through a point $x \not\in (D)$).
\end{itemize}
\begin{property}
An arrow $f \in \Morph(\mathcal M)$ of an affine 2-space has $\varphi(s(f),t(f))$ as inverse if $\mathcal M$ is euclidean, whereas if $\mathcal M$ is hyperbolic then $f$ is not invertible exept if it is an identity arrow.
\end{property}
\begin{proof}
By definition for a euclidean affine 2-space, we have $f = \varphi(t(f),s(f))$ and then $f\circ \varphi(s(f),t(f)) = \varphi(t(f),s(f))\circ \varphi(s(f),t(f)) = \varphi(t(f),t(f)) = \id_{t(f)}$ and $\varphi(s(f),t(f)) \circ f = \varphi(s(f),t(f)) \circ \varphi(t(f),s(f)) = \varphi(s(f),s(f)) = \id_{s(f)}$.\\
For a hyperbolic affine 2-space, let $f$ be an invertible arrow and $f^{-1}$ be its inverse. There exists $(y,a,...,b,x),(x,c,...,d,y)$ such that $\varphi(y,a,...,b,x) = f$ and $\varphi(x,c,...,d,y) = f^{-1}$. We have then $\varphi(y,a,...,b,x) \circ \varphi (x,c,...,d,y) = \varphi(y,a,...,b,x,c,...,d,y) = \varphi(y)$. We have then $(y,a,...,b,x,c,...,d,y) \sim y$ and then $y=x=a=...=b=c=...=d$.
\end{proof}

In order to enlighten the notation, the arrow $\varphi(y,...,x) \in \Morph(\mathcal M)$ of an affine 2-space will be denoted by $\overleftarrow{y...x}$. We have then $\id_x=\overleftarrow{x}$ and $\overleftarrow{z...y} \circ \overleftarrow{y...x} = \overleftarrow{z...y...x}$.\\
Let $\mathcal M$ be an affine 2-space, the category $\mathcal U$ such that $\Obj(\mathcal U)$ is an open submanifold of $\Obj(\mathcal M)$ and $\Morph(\mathcal U) = \varphi \left( \bigsqcup_{n \in \mathbb N^*} \Obj(\mathcal U)^n_{/\mathcal R} \right)$ is called an open affine sub-2-space. By contrast with the generic 2-spaces, it is possible to define easily the union and the intersection of two open affine sub-2-spaces. Let $\mathcal U^1$ and $\mathcal U^2$ two open affine sub-2-spaces of $\mathcal M$. $\mathcal U^1 \cap \mathcal U^2$ and $\mathcal U^1 \cup \mathcal U^2$ are open affine sub-2-spaces defined by
$$ \Obj(\mathcal U^1 \cap \mathcal U^2) = \Obj(\mathcal U^1) \cap \Obj(\mathcal U^2) \qquad \Obj(\mathcal U^1 \cup \mathcal U^2) = \Obj(\mathcal U^1) \cup \Obj(\mathcal U^2) $$
$$ \Morph(\mathcal U^1 \cap \mathcal U^2) = \varphi\left(\bigsqcup_{n \in \mathbb N^*} (\Obj(\mathcal U^1) \cap \Obj(\mathcal U^2))^n_{/\mathcal R} \right) $$
$$ \Morph(\mathcal U^1 \cup \mathcal U^2) = \varphi\left(\bigsqcup_{n \in \mathbb N^*} (\Obj(\mathcal U^1) \cup \Obj(\mathcal U^2))^n_{/\mathcal R} \right) $$
We can note that $\Morph(\mathcal U^1) \cup \Morph(\mathcal U^2) \subsetneq \Morph(\mathcal U^1 \cup \mathcal U^2)$. The composition of arrows belonging to two open affine sub-2-spaces is defined as follows:\\
Let $f \in \Morph(\mathcal U^1)$ and $g \in \Morph(\mathcal U^2)$ with $s(g)=t(f)=y \in \Obj(\mathcal U^1)\cap\Obj(\mathcal U^2)$, $g \circ f = \varphi(z,...,y,...x)$ where $(y,...,x)$ and $(z,...,y)$ are the smaller collections of objets of $\mathcal U^1$ and $\mathcal U^2$ such that $\varphi(y,...,x) = f$ and $\varphi(z,...,y) = g$. We can then define a good open 2-cover of an affine 2-space $\mathcal M$ as being a set of open affine sub-2-spaces $\{\mathcal U^i\}_{i}$ such that $\{\Obj(\mathcal U^i)\}_i$ is a good open cover of $\Obj(\mathcal M)$ (a set of contractible open sets such that $\bigcup_i \Obj(\mathcal U^i) = \Obj(\mathcal M)$). An element $\mathcal U^i$ will be called a 2-chart.\\
In order to enlighten the notation we will simply denote by $M = \Obj(\mathcal M)$ the manifold of objets of an affine 2-space, and by $\{U^i\}_i$ its good open cover.

\begin{example}
Let $(M,G,\cdot)$ be a $G$-space, where $M$ is manifold, $G$ is a Lie group and $\cdot$ is an action of $G$ on $M$. The $G$-space can be viewed as an Euclidean affine space $(\mathcal M,\mathcal R,\varphi)$ with $\Obj \mathcal M = M$ and $\Morph \mathcal M = \{(gG_x,x), x\in M, gG_x \in G/G_x\}$ (where $G_x = \{g \in G, g\cdot x = x\}$ is the stabilizer of $x$). The identity, source and target maps are defined by $\id_x = (G_x,x)$, $s(gG_x,x) = x$ and $t(gG_x,x) = g \cdot x$, the arrow composition being $(hG_{g\cdot x},g\cdot x) \circ (gG_x,x) = (hgG_x,x)$. $x$ and $y$ are linkable if and only if they belong to the same orbit, i.e. $x\mathcal R y \iff x \in G\cdot y$. $\varphi(y,x) = (gG_x,x)$ with $y$ such that $y = g \cdot x$. Remark: an open affine sub-2-space of $\mathcal M$ is not a $G$-space, since $\Morph \mathcal U = \{(gG_x,x); x\in \Obj \mathcal U, gG_x \in G/G_x \text{ such that } g\cdot x \in \Obj \mathcal U \}$.
\end{example}

\section{Categorical principal bundles over affine 2-spaces}
\begin{definition}[Lie crossed module]
A Lie crossed module $\mathcal G$ is the four kinds of data $(G,H,t,\alpha)$ where $G$ and $H$ are Lie groups, $t:H \to G$ and $\alpha:G \to \Aut H$ are homomorphisms such that $t$ is equivariant:
$$ \forall g\in G, \forall h\in H \quad t(\alpha_g(h)) = gt(h)g^{-1} $$
and satisfies the Peiffer identity :
$$ \forall h,h'\in H \quad \alpha_{t(h)}(h') = hh'h^{-1} $$
\end{definition}

\begin{proposition}[A Lie crossed module as a category]
A Lie crossed module is equivalent to a groupoid with $\Obj(\mathcal G) = G$ and $\Morph(\mathcal G) = H \rtimes G$ where the semidirect product (called horizontal composition of arrows) is defined by
$$ (h,g)(h',g') = (h\alpha_g(h'),gg') $$
the identity, source and target maps are defined by
$$ \id_g = (e_H,g) \quad s(h,g)=g \quad t(h,g) = t(h)g $$
and the usual arrow composition (called vertical composition of arrows) is defined by
$$ (h',t(h)g) \circ (h,g) = (h'h,g) $$
\end{proposition}
The Lie crossed modules are the categorical versions of the Lie groups. 

\begin{definition}[Principal 2-bundle over an affine 2-space]
Let $\mathcal M$ be an affine 2-space endowed with a 2-cover $\{\mathcal U^i \}_i$ and $\mathcal G$ be a Lie crossed module. A principal 2-bundle over $\mathcal M$ with structure groupoid $\mathcal G$ consists to a category $\mathcal P$ and a full functor $\pi \in \Funct(\mathcal P,\mathcal M)$ surjective on the objects such that:
\begin{itemize} 
\item $\forall i$, the categories $\mathcal U^i \times \mathcal G$ and $\pi^{-1}(\mathcal U^i)$ are naturally equivalent. We denote by $\phi^i : \mathcal U^i \times \mathcal G \to \pi^{-1}(\mathcal U^i)$ the equivalence (called local trivialisation) and by $\bar \phi^i : \pi^{-1}(\mathcal U^i) \to \mathcal U^i \times \mathcal G$ its weak inverse.
\item The functors $\Pr_1 \bar \phi^i$ and $\pi$ restricted on $\pi^{-1}(\mathcal U^i)$ are equals.
\item The fibration is compatible with the transitive right action of $G$ on itself, i.e. $\forall x \in U^i$, $\forall g,g'\in G$
\begin{equation}
\bar \phi^i \phi^i(x,g)g' = \bar \phi^i \phi^i(x,gg')
\end{equation}
where $(x,g)g'=(x,gg')$.
\end{itemize}
\end{definition}
 We denote by $\kappa^i : \Obj(\mathcal U^i \times \mathcal G) \to \Morph(\mathcal U^i \times \mathcal G)$ the natural equivalence between $\id_{\mathcal U^i \times \mathcal G}$ and $\bar \phi^i \phi^i$, and by $\bar \kappa^i : \Obj(\pi^{-1}(\mathcal U^i)) \to \Morph(\pi^{-1}(\mathcal U^i))$ the natural equivalence between $\id_{\pi^{-1}(\mathcal U^i)}$ and $\phi^i \bar \phi^i$:
\begin{equation}
\forall (x,g) \in U^i \times G, \qquad s(\kappa^i_{xg}) = (x,g) \qquad t(\kappa^i_{xg}) = \bar \phi^i \phi^i (x,g)
\end{equation}
$\forall \overleftarrow{y...x} \in \Morph(\mathcal U^i), \forall h \in H, \forall g \in G$,
\begin{equation}
\label{kfleche}
\kappa^i_{yt(h)g} \circ (\overleftarrow{y...x},h,g) = \bar \phi^i \phi^i(\overleftarrow{y...x},h,g) \circ \kappa^i_{xg}
\end{equation}
\begin{equation}
\forall p \in \Obj(\mathcal P), \qquad s(\bar \kappa^i_p) = p \qquad t(\bar \kappa^i_p) = \phi^i \bar \phi^i (p)
\end{equation}
\begin{equation}
\forall f \in \Morph(\mathcal P), \qquad \bar \kappa^i_{t(f)} \circ f = \phi^i \bar \phi^i (f) \circ \bar \kappa^i_{s(f)}
\end{equation}

\begin{property}
There exists $k^i \in \underline H_{U^i}$ such that :\\
$\forall (x,g) \in U^i \times G$
\begin{equation}
\bar \phi^i \phi^i(x,g) = (x,t(k^i(x))g)
\end{equation}
$ \forall \overleftarrow{y...x}\in \Morph(\mathcal U^i), \forall (h,g) \in H \rtimes G $
\begin{equation}
 \bar \phi^i \phi^i(\overleftarrow{y...x},h,g) = (\overleftarrow{y...x},k^i(y)hk^i(x)^{-1},t(k^i(x))g)
\end{equation}
$ \forall p \in \pi^{-1}(U^i) $
\begin{equation}
 t(\bar \kappa^i_p) = \phi^i(x,t(k^i(x))g_p) 
\end{equation}
with $p = \phi^i(x,g_p)$.
\end{property}

\begin{proof}
Since $s(\kappa^i_{xg}) = (x,g)$, $\exists \mathring k^i_{xg} \in H$ and $\exists (x_n,...,x_1,x)$ such that $\kappa^i_{xg} = (\overleftarrow{x_n...x},\mathring k^i_{xg},g)$. Since $t(\kappa^i_{xg}) = \bar \phi^i \phi^i(x,g)$, we have $x_n=x$ and $t(\mathring k^i_{xg})g = g^i_x$ with $\bar \phi^i \phi^i(x,g) = (x,g^i_x)$. Let $k^i(x)=\mathring k^i_{xe_G}$.  By definition of a 2-bundle we have $\bar \phi^i \phi^i(x,g) = \bar \phi^i \phi^i(x,e_G)g \Rightarrow (x,t(\mathring k^i_{xg})g) = (x,t(k^i(x))g)$ and then $\forall g \in G$, $\mathring k^i_{xg} = k^i(x)$ (modulo an ignored element of $\ker(t)$ without consistent role because it is killed by the target map). We have then $g^i_x = t(k^i(x))g$. This proves the first equality. \\
If $\mathcal M$ is euclidean, then $\overleftarrow{x...x} = \overleftarrow x$. If $\mathcal M$ is spherical, $\overleftarrow{x}$ is the only one arrow with source equal to $x$ (and with target equal to $x$). $\kappa^i$ is a natural equivalence, then $\kappa^i_{xg}$ must be invertible. $\overleftarrow{x...x}$ must be then invertible, now the only invertible arrows of a hyperbolic affine 2-space are the identities, then $\overleftarrow{x...x} = \overleftarrow{x}$ also if $\mathcal M$ is hyperbolic.\\
Let $h \in H$ and $\overleftarrow{y...x} \in \Morph(\mathcal U^i)$. Let $h^i_{y...x} \in H$ be such that $\bar \phi^i \phi^i(\overleftarrow{y...x},h,g) = (\overleftarrow{y...x},h^i_{y...x},g^i_x)$. From equation (\ref{kfleche})  we have
\begin{equation}
(\overleftarrow{y},k^i(y),t(h)g) \circ (\overleftarrow{y...x},h,g) = (\overleftarrow{y...x},h^i_{y...x},g^i_x) \circ (\overleftarrow x, k^i(x),g)
\end{equation}
and then
\begin{equation}
(\overleftarrow{y...x},k^i(y)h,g) = (\overleftarrow{y...x},h^i_{y...x}k^i(x),g)
\end{equation}
We conclude that $h^i_{y...x} = k^i(y)hk^i(x)^{-1}$. This proves the second equality.\\
By definition $t(\bar \kappa^i_p) = \phi^i \bar \phi^i (p) = \phi^i \bar \phi^i \phi^i(x,g_p)$. The third equality comes from $\bar \phi^i \phi^i(x,g_p) = (x,t(k^i(x))g_p)$.
\end{proof}

The natural equivalence is then $\kappa^i_{xg} = (\overleftarrow{x},k^i(x),g)$.

\begin{proposition}[Right actions on a principal 2-bundle]
There is two right actions $R,\bar R : H \rtimes G \to \Funct(\mathcal P,\mathcal P)$ of the Lie crossed module on a principal 2-bundle defined by:\\
$ \forall p \in \Obj(\mathcal P)$ with $p=\phi^i(x,g_p)$
\begin{equation}
R(h,g)p = \phi^i(x,g_pt(h)g)
\end{equation}
$\forall f \in \Morph(\mathcal P)$ with $f=\phi^i(\overleftarrow{y...x},h_f,g_f)$
\begin{equation}
R(h,g)f = \phi^i(\overleftarrow{y...x},h_f,g_ft(h)g)
\end{equation}
$ \forall p \in \Obj(\mathcal P)$ with $\bar \phi^i(p) = (x,\bar g_p)$
\begin{equation}
\bar \phi^i \bar R(h,g)p = (x,\bar g_pt(h)g)
\end{equation}
$\forall f \in \Morph(\mathcal P)$ with $\bar \phi^i(f) = (\overleftarrow{y...x},\bar h_f,\bar g_f)$
\begin{equation}
\bar \phi^i \bar R(h,g) f = (\overleftarrow{y...x},\bar h_f,\bar g_ft(h)g)
\end{equation}
Equivalently, $R$ and $\bar R$ are defined by the following commutative diagrams:
$$ \begin{CD}
\mathcal U^i \times \mathcal G @>{\phi^i}>> \pi^{-1}(\mathcal U^i)  \\
@V{\cdot t(h,g)}VV @VV{R(h,g)}V \\
\mathcal U^i \times \mathcal G @>{\phi^i}>> \pi^{-1}(\mathcal U^i) 
\end{CD}
\qquad \qquad \qquad 
\begin{CD}
 \mathcal U^i \times \mathcal G @<{\bar \phi^i}<< \pi^{-1}(\mathcal U^i) \\
 @A{\cdot t(h,g)}AA @AA{\bar R(h,g)}A \\
\mathcal U^i \times \mathcal G @<{\bar \phi^i}<< \pi^{-1}(\mathcal U^i)
\end{CD}
$$
\end{proposition}

\begin{property}
$\forall (h,g) \in H \rtimes G$, $\bar R(\alpha_{g^{-1}}(h^{-1}),g^{-1}) R(h,g)$ is naturally equivalent to $\id_{\mathcal P}$. Let $\rho(h,g): \Obj(\mathcal P) \to \Morph(\mathcal P)$ be the associated natural equivalence. $\forall p \in \Obj(\mathcal P)$, $s(\rho_p(h,g)) = p$ and $\bar \phi^i t(\rho_p(h,g)) = (x,t(k^i(x))g_p)$ with $p=\phi^i(x,g_p)$.\\
$\forall (h,g) \in H \rtimes G$, $R(h,g)\bar R(\alpha_{g^{-1}}(h^{-1}),g^{-1})$ is naturally equivalent to $\id_{\mathcal P}$. Let $\bar \rho(h,g): \Obj(\mathcal P) \to \Morph(\mathcal P)$ be the associated natural equivalence. $\forall p \in \Obj(\mathcal P)$, $s(\bar \rho_p(h,g)) = \phi^i(x,t(k^i(x))\bar g_p)$ with $\bar \phi^i(p)=(x,\bar g_p)$ and $t(\bar \rho_p(h,g)) = p$.
\end{property}

\begin{proof}
The natural equivalences follow from the following diagram:
\begin{eqnarray*}
\mathcal U^i \times \mathcal G \quad & \xleftarrow[\quad \bar \phi^i \quad]{\displaystyle \xrightarrow{\quad \phi^i \quad}} & \quad \pi^{-1}(\mathcal U^i) \\
{\scriptstyle \cdot t(\alpha_{g^{-1}}(h^{-1}),g^{-1})} \uparrow \downarrow {\scriptstyle \cdot t(h,g)} & & {\scriptstyle R(h,g)} \downarrow \uparrow {\scriptstyle \bar R(\alpha_{g^{-1}}(h^{-1}),g^{-1})} \\
\mathcal U^i \times \mathcal G \quad & \xleftarrow[\quad \bar \phi^i \quad]{\displaystyle \xrightarrow{\quad \phi^i \quad}} & \quad \pi^{-1}(\mathcal U^i)
\end{eqnarray*}
where each double arrow is naturally equivalent to an identity.\\
By definition, 
$$ t(\rho_p(h,g)) = \bar R(\alpha_{g^{-1}}(h^{-1}),g^{-1}) R(h,g) p = \bar R(\alpha_{g^{-1}}(h^{-1}),g^{-1}) \phi^i(x,g_pt(h)g)$$
with $p=\phi^i(x,g_p)$. We have then $\bar \phi^i t(\rho_p(h,g)) = \bar \phi^i\bar R(\alpha_{g^{-1}}(h^{-1}),g^{-1}) \phi^i(x,g_pt(h)g)$. Since $\bar \phi^i \phi^i (x,g_pt(h)g) = (x,t(k^i(x))g_pt(h)g)$ we have 
$$\bar \phi^i t(\rho_p(h,g)) = (x,t(k^i(x)) g_pt(h)gg^{-1}t(h^{-1}))$$
By definition, $s(\bar \rho_p(h,g)) = R(h,g)\bar R(\alpha_{g^{-1}}(h^{-1}),g^{-1})p$. Since 
$$ \bar \phi^i \bar R(\alpha_{g^{-1}}(h^{-1}),g^{-1}) p =(x,\bar g_pg^{-1}t(h^{-1})) $$ 
with $\bar \phi^i(p) = (x,\bar g_p)$, we have 
$$ \phi^i \bar \phi^i \bar R(\alpha_{g^{-1}}(h^{-1}),g^{-1}) p = t(\bar \kappa^i_{\phi^i(x,\bar g_pg^{-1}t(h^{-1}))}) = \phi^i(x,t(k^i(x)) \bar g_pg^{-1}t(h^{-1}))$$
And by definition of $R$, we have 
$$R(h,g)\bar R(\alpha_{g^{-1}}(h^{-1}),g^{-1})p = \phi^i(x,t(k^i(x)) \bar g_pg^{-1}t(h^{-1})t(h)g)$$
\end{proof}

\begin{definition}[$G$-transition functions]
We define the $G$-transition functions of a principal 2-bundle as being $g^{ij} \in \underline G_{U^i \cap U^j}$ such that $\forall x \in U^i \cap U^j$, $\forall g\in G$
\begin{equation}
\bar \phi^i \phi^j (x,g) = (x,g^{ij}(x)g)
\end{equation}
\end{definition}

\begin{property}
The $G$-transition functions satisfy
\begin{equation}
g^{ii}(x) = t(k^i(x)) \qquad g^{ji}(x) = t(k^j(x)) g^{ij}(x)^{-1} t(k^i(x))
\end{equation}
\end{property}

\begin{proof}
$(x,g^{ii}(x)) = \bar \phi^i \phi^i(x,e_G) = (x,t(k^i(x)))$.
$$ \bar \phi^i \phi^j \bar \phi^j \phi^i(x,e_G) = \bar \phi^i \phi^j(x,g^{ji}(x)) = (x,g^{ij}(x)g^{ji}(x))$$
But we have also 
$$ \bar \phi^i \phi^j \bar \phi^j \phi^i(x,e_G) = \bar \phi^i t(\bar \kappa^j_p) $$
 with $p= \phi^i(x,e_G)$. Moreover $t(\bar \kappa^j_p) = \phi^j(x,t(k^j(x))g_p)$ with $ p = \phi^j(x,g_p)$. 
$$\phi^i(x,e_G) = \phi^j(x,g_p) \Rightarrow \bar \phi^i \phi^i(x,e_G) = \bar \phi^i \phi^j(x,g_p) $$ 
and then 
$$ (x,t(k^i(x))) = (x,g^{ij}(x)g_p) \Rightarrow g_p = g^{ij}(x)^{-1} t(k^i(x))$$
Finally 
\begin{eqnarray*}
\bar \phi^i \phi^j \bar \phi^j \phi^i (x,e_G) & = & \bar \phi^i \phi^j (x,t(k^j(x))g^{ij}(x)^{-1}t(k^i(x))) \\
& = & (x,g^{ij}(x)t(k^j(x))g^{ij}(x)^{-1}t(k^i(x)))
\end{eqnarray*}
 We conclude that $g^{ji}(x)=t(k^j(x)) g^{ij}(x)^{-1}t(k^i(x))$.
\end{proof}

\begin{definition}[$H$-transition functions]
We define the $H$-transition functions of a principal 2-bundle as being $h^{ij} \in \underline H_{(U^i \cap U^j)^2}$ such that $\forall \overleftarrow{y...x} \in \Morph(\mathcal U^i \cap \mathcal U^j)$, $\forall (h,g) \in H \rtimes G$
\begin{eqnarray}
\bar \phi^i \phi^j(\overleftarrow{y...x},h,g) & = & \left(\overleftarrow{y...x},(h^{ij}(y,x),g^{ij}(x))(h,g)\right) \\
& = & (\overleftarrow{y...x},h^{ij}(y,x)\alpha_{g^{ij}(x)}(h),g^{ij}(x)g)
\end{eqnarray}
\end{definition}
The fact that $h^{ij}(\overleftarrow{y...x})$ depends only from the source and the target of $\overleftarrow{y...x}$ (even if the affine 2-space is hyperbolic) is a consequence of the following property:
\begin{property}
The $G$-transition functions and the $H$-transition functions are related by $\forall x,y \in U^i \cap U^j$ such that $\exists \overleftarrow{y...x}\in \Morph(\mathcal U^i \cap \mathcal U^j)$,
\begin{equation}
t(h^{ij}(y,x)) g^{ij}(x) = g^{ij}(y)
\end{equation}
\end{property}

\begin{proof}
We have $t(\bar \phi^i \phi^j (\overleftarrow{y...x},e_H,e_G)) = \bar \phi^i \phi^j(y,e_G) = (y,g^{ij}(y))$. But we have also $t(\bar \phi^i \phi^j (\overleftarrow{y...x},e_H,e_G)) = t(\overleftarrow{y...x},h^{ij}(y,x),g^{ij}(x)) = (y, t(h^{ij}(y,x))g^{ij}(x))$.
\end{proof}

In fact, we could have $h^{ij}(\overleftarrow{yz_n...z_1x}) = h^{ij}(y,x) \zeta^{ij}(z_n,...,z_1)$ with $\zeta^{ij} \in \ker(t)$. Since $\zeta^{ij}$ presents no consistent information (because it is killed by the target map), for the sake of simplicity we consider that $\zeta^{ij} = e_H$. For the same reason we consider that $h^{ij}(x,x) = e_H$, the $H$-transition functions are then trivial if the affine 2-space is spheric.

\begin{property}
The $H$-transition functions satisfy
\begin{equation}
h^{ii}(y,x) = k^i(y)k^i(x)^{-1}
\end{equation}
\begin{equation}
h^{ji}(y,x) = k^j(y) \alpha_{g^{ij}(x)^{-1}}(h^{ij}(y,x)^{-1} k^i(y)k^i(x)^{-1}) k^j(x)^{-1} 
\end{equation}
\end{property}

\begin{proof}
$$ (\overleftarrow{y...x},h^{ii}(y,x),g^{ii}(x)) = \bar \phi^i \phi^i(\overleftarrow{y...x},e_H,e_G) = (x,k^i(y)k^i(x)^{-1},t(k^i(x)))$$
We have 
$$\bar \phi^i \phi^j \bar \phi^j \phi^i(\overleftarrow{y...x},e_H,e_G) = (\overleftarrow{y...x},h^{ij}(y,x)\alpha_{g^{ij}(x)}(h^{ji}(y,x)),g^{ij}(x)g^{ji}(x))$$
 Moreover we have 
$$ \bar \phi^i \phi^j \bar \phi^j \phi^i(\overleftarrow{y...x},e_H,e_G) = \bar \phi^i \phi^j \bar \phi^j(f)$$ 
with $f = \phi^i(\overleftarrow{y...x},e_H,e_G)$. But we have 
$$ \bar \phi^i \phi^j \bar \phi^j(f) = \bar \phi^i(\bar \kappa^j_q \circ f \circ \bar \kappa^{j-1}_p)$$
 with $p = \phi^i(x,e_G)$ and $q=\phi^i(y,e_G)$. 
$$ \bar \phi^i(\bar \kappa^j_q \circ f \circ \bar \kappa^{j-1}_p) = \bar \phi^i \phi^j(\overleftarrow{y...x},k^j(y) h_p k^j(x)^{-1}, t(k^j(x))g_p)$$
 with $f=\phi^j(\overleftarrow{y....x},h_p,g_p)$. 
$$\phi^i(\overleftarrow{y...x},e_H,e_G) = \phi^j(\overleftarrow{y...x},h_p,g_p) \Rightarrow \bar \phi^i \phi^i(\overleftarrow{y...x},e_H,e_G) = \bar \phi^i \phi^j(\overleftarrow{y...x},h_p,g_p)$$
and then 
$$ (\overleftarrow{y...x},k^i(y)k^i(x)^{-1},t(k^i(x))) = (\overleftarrow{y...x},h^{ij}(y,x)\alpha_{g^{ij}(x)}(h_p),g^{ij}(x)g_p)$$
 It follows that $h_p = \alpha_{g^{ij}(x)^{-1}}(h^{ij}(y,x)^{-1}k^i(y)k^i(x)^{-1})$. Finally we can identify $h^{ij}(y,x)\alpha_{g^{ij}(x)}(h^{ji}(y,x))$ with $h^{ij}(y,x)\alpha_{g^{ij}(x)}(k^j(y)h_pk^j(x)^{-1})$:
\begin{eqnarray*}
& & h^{ij}(y,x) \alpha_{g^{ij}(x)}(h^{ji}(y,x)) \nonumber \\
& &  = h^{ij}(y,x) \alpha_{g^{ij}(x)}(k^j(y) \alpha_{g^{ij}(x)^{-1}}(h^{ij}(y,x)^{-1} k^i(y)k^i(x)^{-1}) k^j(x)^{-1})
\end{eqnarray*}
\end{proof}

\begin{property}
We consider the sub-2-bundle $\pi^{-1}(\mathcal U^i \cap \mathcal U^j \cap \mathcal U^k)$ at the intersection of three 2-charts. The functors $\bar \phi^i \phi^j \bar \phi^j \phi^k$ and $\bar \phi^i \phi^k$ (restricted on $\mathcal U^i \cap \mathcal U^j \cap \mathcal U^k$) are naturally equivalents.
\end{property}

\begin{proof}
This follows from the following commutative diagram:
$$ \begin{CD}
\mathcal U^{ijk} \times \mathcal G @>{\phi^k}>> \pi^{-1}(\mathcal U^{ijk}) @>{\phi^j \bar \phi^j}>> \pi^{-1}(\mathcal U^{ijk}) @>{\bar \phi^i}>>  \mathcal U^{ijk} \times \mathcal G \\
& @| @V{\kappa^i}V{{\displaystyle \uparrow} \bar \kappa^i}V @| \\
\mathcal U^{ijk} \times \mathcal G @>>{\phi^k}> \pi^{-1}(\mathcal U^{ijk}) @>>{\id_{\pi^{-1}(\mathcal U^j)}}> \pi^{-1}(\mathcal U^{ijk}) @>>{\bar \phi^i}>  \mathcal U^{ijk} \times \mathcal G
\end{CD}
$$
where $\mathcal U^{ijk} = \mathcal U^i \cap \mathcal U^j \cap \mathcal U^k$.
\end{proof}

Let $\breve h^{ijk} : \Obj(\mathcal U^i \cap \mathcal U^j \cap \mathcal U^k \times \mathcal G) \to \Morph(\mathcal U^i \cap \mathcal U^j \cap \mathcal U^k \times \mathcal G)$ be the associated natural equivalence:
\begin{equation}
s(\breve h^{ijk}_{xg}) = \bar \phi^i \phi^j \bar \phi^j \phi^k(x,g) = (x,g^{ij}(x)g^{jk}(x)g)
\end{equation}
\begin{equation}
t(\breve h^{ijk}_{xg}) = \bar \phi^i \phi^k(x,g) = (x,g^{ik}(x)g)
\end{equation}
\begin{equation}
\breve h^{ijk}_{yt(h)g} \circ \bar \phi^i \phi^j \bar \phi^j \phi^k(\overleftarrow{y...x},h,g) = \bar \phi^i \phi^k(\overleftarrow{y...x},h,g) \circ \breve h^{ijk}_{xg}
\end{equation}
Let $h^{ijk}(x) \in H$ be such that
\begin{equation}
\breve h^{ijk}_{xe_G} = (\overleftarrow x,h^{ijk}(x),g^{ij}(x)g^{jk}(x))
\end{equation}
Since $t(\breve h^{ijk}_{xe_G}) = (x,g^{ik}(x))$ we have $g^{ik}(x) = t(h^{ijk}(x))g^{ij}(x)g^{jk}(x)$.
\begin{definition}[2-transition functions]
We define the 2-transition functions of a principal 2-bundle as being $h^{ijk} \in \underline H_{U^i \cap U^j \cap U^k}$ such that $\forall x \in U^i \cap U^j \cap U^k$
\begin{equation}
t(h^{ijk}(x)) g^{ij}(x)g^{jk}(x) = g^{ik}(x)
\end{equation}
\end{definition}
The 2-transition functions measure then the obstruction to lift $\mathcal P$ as an usual principal bundle, since they charaterizes the failure of the cocycle relation for the $G$-transition functions. We can remark that 
\begin{eqnarray} 
h^{iji}(x) & = & \alpha_{g^{ij}(x)}(k^j(x)^{-1}) \\
h^{iij}(x) & = & k^i(x)^{-1} \\
h^{ijj}(x) & = & \alpha_{g^{ij}(x)}(k^j(x)^{-1})
\end{eqnarray}

\begin{property}
The natural equivalence $\breve h^{ijk}$ for some $g\in G$ is
\begin{equation}
\breve h^{ijk}_{xg} = (\overleftarrow x,h^{ijk}(x), g^{ij}(x)g^{jk}(x)g)
\end{equation}
\end{property}

\begin{proof}
Let $\mathring h^{ijk}_{xg} \in H$ be such that $\breve h^{ijk}_{xg} = (\overleftarrow x,\mathring h^{ijk}_{xg}, g^{ij}(x)g^{jk}(x)g)$. We have then $t(\mathring h^{ijk}_{xg})g^{ij}(x)g^{jk}(x)g = g^{ik}(x)g$ and $g^{ik}(x) = t(h^{ijk}(x))g^{ij}(x)g^{jk}(x)$.
\end{proof}

\begin{property}
The 2-transition functions measure also the failure of the cocycle relation for the $H$-transition functions in the following sense:
\begin{equation}
\label{Hsteq}
h^{ij}(y,x)\alpha_{g^{ij}(x)}(h^{jk}(y,x)) = h^{ijk}(y)^{-1} h^{ik}(y,x) h^{ijk}(x)
\end{equation}
\end{property}

\begin{proof}
By using the definition of $\breve h^{ijk}_{xe_G}$ and the expressions of $\bar \phi^i \phi^j \bar \phi^j \phi^k(\overleftarrow{y...x},e_H,e_G)$ and of $\bar \phi^i \phi^k(\overleftarrow{y...x},e_H,e_G)$, we find
\begin{eqnarray*}
& & (\overleftarrow y, h^{ijk}(y),g^{ij}(y)g^{jk}(y)) \circ (\overleftarrow{y...x},h^{ij}(y,x) \alpha_{g^{ij}(x)}(h^{jk}(y,x)),g^{ij}(x)g^{jk}(x)) \nonumber \\
& & \qquad = (\overleftarrow{y...x},h^{ik}(y,x),g^{ik}(x)) \circ (\overleftarrow x,h^{ijk}(x),g^{ij}(x)g^{jk}(x))
\end{eqnarray*}
By composing the arrows, we find
\begin{eqnarray*}
& & (\overleftarrow{y...x},h^{ijk}(y) h^{ij}(y,x)\alpha_{g^{ij}(x)}(h^{jk}(y,x)),g^{ij}(x)g^{jk}(x)) \nonumber \\
& & \qquad = (\overleftarrow{y...x},h^{ik}(y,x) h^{ijk}(x),g^{ij}(x)g^{jk}(x))
\end{eqnarray*}
\end{proof}

Since $\breve h^{ijk}$ is a natural equivalence, it has an inverse $\breve h^{ijk-1}$ such that $\forall x \in U^i \cap U^j \cap U^k$ and $\forall g\in G$, $\breve h^{ijk-1}_{xg} \circ \breve h^{ijk}_{xg} = (\overleftarrow x, e_H, g^{ij}(x)g^{jk}(x)g)$. It is clear that $\breve h^{ijk-1}_{xg} = (\overleftarrow x, h^{ijk}(x)^{-1},g^{ik}(x)g)$ where $h^{ijk}(x)^{-1}$ is the inverse of $h^{ijk}(x)$ in the group law sense.

\begin{property}
The 2-transition functions $h^{ijk}$ can be viewed as the trivializations of the natural equivalence $\bar \kappa^j$ on $\mathcal U^i \cap \mathcal U^j \cap \mathcal U^k$ since we have
\begin{equation}
(\overleftarrow{x}, h^{ijk}(x)^{-1},g^{ik}(x)) = \bar \phi^i(\bar \kappa^j_{\phi^k(x,e_G)})
\end{equation}
\end{property} 

\begin{proof}
By definition of $\bar \kappa^j$ we have 
$$\phi^j \bar \phi^j \phi^k(\overleftarrow{y...x},e_H,e_G) = \bar \kappa^j_{\phi^k(y,e_G)} \circ \phi^k(\overleftarrow{y...x},e_H,e_G) \circ \bar \kappa^{j-1}_{\phi^k(x,e_G)}$$
 We have then 
$$\bar \phi^i \phi^j \bar \phi^j \phi^k(\overleftarrow{y...x},e_H,e_G) = \bar \phi^i(\bar \kappa^j_{\phi^k(y,e_G)}) \circ \bar \phi^i \phi^k(\overleftarrow{y...x},e_H,e_G) \circ \bar \phi^i(\bar \kappa^{j-1}_{\phi^k(x,e_G)})$$
 By using the definition of $\breve h^{ijk}$ we have 
$$\breve h^{ijk}_{ye_G} \circ \bar \phi^i(\bar \kappa^j_{\phi^k(y,e_G)}) \circ \bar \phi^i \phi^k(\overleftarrow{y...x},e_H,e_G) \circ \bar \phi^i(\bar \kappa^{j-1}_{\phi^k(x,e_G)}) = \bar \phi^i \phi^k(\overleftarrow{y...x},e_H,e_G) \circ \breve h^{ijk}_{xe_G}$$
 We can conclude that 
$$ \breve h^{ijk}_{xe_G} \circ \bar \phi^i(\bar \kappa^j_{\phi^k(x,e_G)}) = \id_{\bar \phi^i \phi^k(x,e_G)} \Rightarrow \breve h^{ijk-1}_{xe_G} = \bar\phi^i(\bar \kappa^j_{\phi^k(x,e_G)})$$
\end{proof}

\begin{property}
The 2-transition functions obey to the generalized cocyle relation:$\forall x \in U^i\cap U^j \cap U^k \cap U^l$
\begin{equation}
h^{ijl}(x) \alpha_{g^{ij}(x)}(h^{jkl}(x)) = h^{ikl}(x)h^{ijk}(x)
\end{equation}
\end{property}

\begin{proof}
This follows from the definition of the 2-transition functions:
\begin{eqnarray*}
t(h^{ijl}(x))(t(\alpha_{g^{ij}(x)}(h^{jkl}(x))) g^{ij}(x)g^{jk}(x)g^{kl}(x) & = & t(h^{ijl}(x)) g^{ij}(x)t(h^{jkl}(x))g^{jk}(x)g^{kl}(x) \\
& = & t(h^{ijl}(x))g^{ij}(x)g^{jl}(x)=g^{il}(x)
\end{eqnarray*}
and 
$$t(h^{ikl}(x))t(h^{ijk}(x))g^{ij}(x)g^{jk}(x)g^{kl}(x) = t(h^{ikl}(x))g^{ik}(x)g^{kl}(x)=g^{il}(x)$$
\end{proof}

Our definition of a principal 2-bundle coincides with the definition of Baez \textit{etal} \cite{baez1,baez2,baez3,wockel,soncini} for the spherical affine 2-spaces, but our theory is more general since it can be applied with non-trivial base categories (with not only identity arrows) as euclidean and hyperbolic affine 2-spaces. Since $H$-transition functions are trivial with a spherical affine 2-space ($h^{ij}(y,x) = e_H$, $\forall x \mathcal R y \iff x=y$), these local data of the 2-bundles are absent from the theory of Baez \textit{etal}. Because the non-abelian bundle gerbes \cite{laurent,breen,kalkkinen2} are weakly equivalent to 2-bundles \cite{nikolaus} the same remarks can be applied in the comparison of our definition with the constructions of non-abelian bundle gerbes or twisted bundles. Nevertheless, the non-abelian bundle gerbes present a kind of $H$-transition functions obeying to a structure equation similar to equation \ref{Hsteq} (see \cite{aschieri2}). But in that case, the $H$-transition functions are not associated with arrows in a base category but with points of the manifold $Y \times_M Y \to M$ where $M$ is the base manifold and $Y \to M$ is a fibre bundle (the non-abelian gerbe construction consists to three floor local $H$-principal bundles over $Y \times _M Y$ and $Y \times_M Y \times_M Y$ \cite{aschieri2} where all entities are usual manifolds and not explicitely categories as in our construction). Categorical bundles over pathspaces \cite{cattaneo,chatterjee,chatterjee2} are defined over non-trivial categories, i.e. over pathspaces of manifolds viewed as categories. Such categories are not affine 2-spaces and then the two constructions are completely separated. Since a bundle over the pathspace of a manifold is built from an usual principal bundle with connection over this manifold, it presents a trivial 2-transition functions ($h^{ijk}(x) = e_H$). To summarize, in term of local data defining a categorical bundle, our construction seems the more general because it presents possibly non-trivial $H$-transition functions (associated with arrows of a base category) and 2-transition functions.\\

\begin{example}
Let $(\xi_a)_{a=1,...,n}$ be the canonical basis of $\mathbb C^n$ endowed with the usual inner product. We denotes by $GL(n,\mathbb C)$ and $U(n)$ the Lie groups of invertible and unitary matrices expressed in this canonical basis. Let $U(m)$ be the subgroup of $U(n)$ ($m<n$) of unitary matrices of $\mathbb C^m$ generated by $(\xi_a)_{a=1,...,m}$. We call density matrix of $\mathbb C^n$ a $n \times n$ matrix $\rho$ such that $\rho^\dagger = \rho$, $\rho \geq 0$, and $\tr \rho= 1$. Let $\sigma$ be a diagonal density matrix such that $\forall a,b$, $a\not=b$, $\sigma_{aa} \not= \sigma_{bb}$; $\forall a>m$, $\sigma_{aa} = 0$, and $\sum_{a=1}^m \sigma_{aa} = 1$. Let $M$ be the manifold of density matrices of $\mathbb C^n$ which are isospectral to $\sigma$. For $\rho \in M$, let $F_\rho = \{f \in GL(n,\mathbb C), \text{such that } f\rho f^\dagger \text{ is isospectral to } \rho\}$. Let $H_\rho$ be the stabilizer of $\rho$ for the conjugation, i.e. $H_\rho=\{h\in GL(n,\mathbb C), h\rho h^\dagger = \rho\}$. $H_\rho \subset F_\rho$ and $F_\rho / H_\rho \simeq U(n) / U(n)_\rho$ (where $U(n)_\rho$ is the stabilizer of $\rho$ for the adjoint action, i.e. $U(n)_\rho = \{h \in U(n), h\rho h^{-1}=\rho\}$). For all $\rho$, $H_\rho$ is isomorphic to $H \equiv H_\sigma$. Let $\mathcal M$ be the totally linkable euclidean affine 2-space defined by $\Obj(\mathcal M) = M$, $\Morph(\mathcal M) = \{([f],\rho), \rho \in M, [f]\in F_\rho/H_\rho\}$, with $s([f],\rho) = \rho$, $t([f],\rho) = f\rho f^\dagger$ ($f$ being an element of the coset $[f]$), $\id_\rho = ([\id_{\mathbb C^n}],\rho)$, and $([f'],f\rho f^\dagger) \circ ([f],\rho) = ([f'f],\rho)$.\\
We call a purification of $\rho \in M$, a matrix $W \in \mathfrak M_{n\times n}(\mathbb C)$ such that $\rho = WW^\dagger$. $W$ is not unique, and one of the purifications of $\rho$ is $\sqrt \rho$. Let $\{U^i\}_i$ be a good open cover of $M$ such that $\forall \rho,\rho' \in U^i$, $\Ran \rho \cap \ker \rho' = \ker \rho \cap \Ran \rho' = \{0\}$. For all $\rho \in U^i$, we choose $(\chi^i_{\rho a})_{a=1,...,m}$ an orthonormal basis of $\Ran \rho$ (continuous in norm with respect to $\rho$). Let $Z^i_\rho \in \mathfrak M_{n \times m}(\mathbb C)$ be the matrix representing $(\chi^i_{\rho a})_{a=1,...,m}$  in the canonical basis, i.e. $Z^i_{\rho ab} = \langle \xi_a|\chi^i_{\rho b} \rangle$. Note that $Z^{i\dagger}_{\rho} Z^i_\rho = \id_{\mathbb C^m}$ and $Z^i_\rho Z^{i \dagger}_\rho = P_{\Ran \rho}$ (orthogonal projection onto $\Ran \rho$). By construction, $\forall \rho \in U^i$, $\exists f^i_\rho \in F_{Z^i_{\rho} \sigma Z^{i\dagger}_{\rho}}$ such that $\rho = f^i_\rho Z^i_{\rho} \sigma Z^{i\dagger}_{\rho} f^{i\dagger}_{\rho}$. Let $G=U(m)$, $g \in U(m)$ defines a basis change of $\Ran \rho$ by its right action : $Z^i_\rho g$.\\
Note that $\sqrt \sigma G \sqrt \sigma^{-1} \subset H$ (where $\sigma^{-1}$ denotes the pseudoinverse of $\sigma$, i.e. $\forall i>m$, $(\sigma^{-1})_{ii} = 0$ and $\forall i\leq m$, $(\sigma^{-1})_{ii} = (\sigma_{ii})^{-1}$). Indeed, $\sqrt \sigma g \sqrt \sigma^{-1} \sigma (\sqrt \sigma g \sqrt \sigma^{-1})^\dagger = \sqrt \sigma g \sqrt \sigma^{-1}\sigma \sqrt \sigma^{-1} g^{-1} \sqrt{\sigma} = \sigma$ ($\forall g \in U(m)$). Let $\mathcal G = (G,H,t,\alpha)$ be the Lie crossed module defined by $t(h) = \sqrt \sigma^{-1} h \sqrt \sigma$ and $\alpha_g(h) = \sqrt \sigma g \sqrt \sigma^{-1} h \sqrt \sigma g^{-1} \sqrt \sigma^{-1}$.\\
The purification of the density matrices defines a principal 2-bundle $\mathcal P$ over $\mathcal M$ with structure groupoid $\mathcal G$, where $\Obj(\mathcal P)$ is the set of the purifications of $M$, $\Morph \mathcal P = \{(f,W), W \in \Obj(\mathcal P), f \in F_{WW^\dagger}\}$ (with $s(f,W)=W$, $t(f,W)=fW$, $\id_W = (\id_{\mathbb C^n},W)$, $(f',fW)\circ(f,W) = (f'f,W)$). The projection functor $\pi: \mathcal P \to \mathcal M$ is defined by $\pi(W)= WW^\dagger$ and $\pi(f,W) = ([f],WW^\dagger)$ ($[f] \in F_{WW^\dagger}/H_{WW^\dagger}$). The local trivialisations of $\mathcal P$, $\phi^i : \mathcal U^i \times \mathcal G \to \mathcal P_{|\mathcal U^i}$, are defined by:
$$ 
\begin{array}{ccc}
\begin{CD}
(\rho,g) \\ @V{([f],\rho,h,g)}VV \\ (f\rho f^\dagger,\sqrt \sigma^{-1} h \sqrt \sigma g)
\end{CD}
&
\overset{\phi^i}{\Longrightarrow}
&
\begin{CD}
f^i_\rho Z^i_\rho \sqrt \sigma g \\ @VV{(\mathring f f^i_\rho Z^i_\rho h Z^{i\dagger}_\rho f^{i-1}_\rho,f^i_\rho Z^i_\rho g)}V \\ f^i_{f\rho f^\dagger} Z^i_{f\rho f^\dagger} h \sqrt \sigma g
\end{CD}
\end{array}
$$
where $\mathring f \in [f] \cap U(n)$. We note that because of $f^i_{f\rho f^\dagger} = \mathring f f^i_\rho \mathring f^{-1}$ and $Z^i_{f\rho f^\dagger} = \mathring f Z^i_\rho$, it follows that $f^i_{f\rho f^\dagger} Z^i_{f\rho f^\dagger} h \sqrt \sigma g = \mathring f f^i_\rho Z^i_\rho h \sqrt \sigma g$. The inverse trivialisations $\bar \phi^i: \mathcal P_{\mathcal U^i} \to \mathcal U^i \times \mathcal G$ are defined by:
$$ 
\begin{array}{ccc}
\begin{CD}
W \\ @V{(f,W)}VV \\ fW
\end{CD}
&
\overset{\bar \phi^i}{\Longrightarrow}
&
\begin{CD}
(WW^\dagger,\sqrt \sigma^{-1} Z^{i\dagger}_{WW^\dagger} f^{i\dagger}_{WW^\dagger} W) \\ @VV{([f],WW^\dagger,Z^{i\dagger}_{WW^\dagger}f^{i\dagger}_{WW^\dagger} \mathring f^{-1} f f^{i\dagger-1}_{WW^\dagger} Z^i_{WW^\dagger},\sqrt \sigma^{-1} Z^{i\dagger}_{WW^\dagger} f^{i\dagger}_{WW^\dagger}W)}V \\ (fWW^\dagger f,\sqrt \sigma^{-1} Z^{i\dagger}_{fWW^\dagger f} f^{i\dagger}_{fWW^\dagger f^\dagger} f W)
\end{CD}
\end{array}
$$
We note that $\sqrt \sigma^{-1} Z^{i\dagger}_{fWW^\dagger f} f^{i\dagger}_{fWW^\dagger f^\dagger} f W = \sqrt \sigma^{-1} Z^{i\dagger}_{WW^\dagger} f^{i\dagger}_{WW^\dagger} \mathring f^{-1} f W$.\\
By using the expressions of these trivializations, the local data of $\mathcal P$ are $k^i(\rho) = Z^{i\dagger}_\rho f^{i\dagger}_\rho f^i_\rho Z^i_\rho$, $g^{ij}(\rho) = Z^{i\dagger}_\rho f^{i\dagger}_\rho f^j_\rho Z^j_\rho$ , $h^{ij}(f,\rho) = \sqrt \sigma \mathring f g^{ij}(\rho) \mathring f^{-1} g^{ij}(\rho)^{-1} \sqrt \sigma^{-1}$, and\\ $h^{ijk}(\rho) = \sqrt \sigma Z^{i\dagger}_\rho f^{i\dagger}_\rho f^k_\rho P_{\Ran \rho} f^{k-1}_\rho f^{j\dagger-1}_\rho P_{\Ran \rho} f^{j-1}_\rho f^{i\dagger-1}_\rho Z^i_\rho \sqrt \sigma^{-1}$.\\
We note that in the case where $m=1$ (i.e. where $M$ is the space of pure states, $\rho$ is a projection, $\rho^2=\rho$), $f^i_\rho \in U(1)$ and $\mathcal P$ is trivial in the sense where $h^{ijk}(\rho) = 1$ and $\Obj \mathcal P$ is the Berry-Simon $U(1)$-bundle \cite{simon}.
\end{example}

\section{2-connections}
The definition of a connective structure on a principal 2-bundle over an affine 2-space needs to introduce the ``Lie algebra like'' of a Lie crossed bundle. After this, before to consider the generic 2-connections, it is instructive to study the case of a trivial 2-bundle.
\subsection{Differential Lie crossed module}
\begin{definition}[Differential Lie crossed module]
Let $\mathcal G = (G,H,t,\alpha)$ be a Lie crossed module. The differential Lie crossed module associated with $\mathcal G$ is the four kinds of data $(\mathfrak g,\mathfrak h, t^{Lie},\alpha^{Lie})$ where $\mathfrak g$ and $\mathfrak h$ are the Lie algebras of $G$ and $H$, and $t^{Lie}:\mathfrak h \to \mathfrak g$ and $\alpha^{Lie} : \mathfrak g \to \mathrm{Der}(\mathfrak h)$ are the maps induced by $t$ and $\alpha$ in the Lie algebras, so
$$ \forall X \in \mathfrak g, \forall Y \in \mathfrak h, \quad t^{Lie}(\alpha^{Lie}_X(Y)) = [X,t^{Lie}(Y)] $$
and
$$ \forall Y,Y' \in \mathfrak h, \quad \alpha^{Lie}_{t^{Lie}(Y)}(Y')=[Y,Y'] $$
\end{definition}

The semi-direct product of groups $H \rtimes G$ induces a semi-direct sum of Lie algebras $\mathfrak h \soplus \mathfrak g$ defined as being $\mathfrak h \oplus \mathfrak g$ (the exterior direct sum being between the vector spaces without the algebra structures) endowed with the Lie braket $[.,.]_s$ such that
$$ \forall X,X' \in \mathfrak g, \quad [X,X']_s = [X,X']_{\mathfrak g} \in \mathfrak g $$
$$ \forall Y,Y' \in \mathfrak h, \quad [Y,Y']_s = [Y,Y']_{\mathfrak h} \in \mathfrak h $$
$$ \forall X \in \mathfrak g,\forall Y \in \mathfrak h, \quad [X,Y]_s = - [Y,X]_s = \alpha^{Lie}_X(Y) \in \mathfrak h $$
To simplify the notation, we denote all the Lie brakets by $[.,.]$ without subsript.\\
In the following, we denote by $\pi^{\mathfrak g} : \mathfrak h \soplus \mathfrak g \to \mathfrak g$ the projection induced by the canonical projection $\mathfrak h \oplus \mathfrak g \to \mathfrak g$ defined by the exterior direct sum of vector spaces.\\

Moreover, we will denote the adjoint representation of $H \rtimes G$ on $\mathfrak h \soplus \mathfrak g$ by
\begin{equation}
\forall h\in H, \forall g \in G, \forall X \in \mathfrak h \soplus \mathfrak g, \quad \Ad(h,g)X = hgXg^{-1}h^{-1}
\end{equation}
This notation is in accordance with the semi-direct product since
\begin{eqnarray}
\Ad(h_2,g_2) \Ad(h_1,g_1) X & = & h_2g_2h_1g_1 X g_1^{-1} h_1^{-1} g_2^{-1} h_2^{-1} \\
& = & (h_2 g_2 h_1 g_2^{-1})g_2 g_1 X g_1^{-1} g_2^{-1} (g_2 h_1^{-1} g_2^{-1} h_2^{-1}) \\
& = & \Ad(h_2 \alpha_{g_2}(h_1),g_2g_1) X
\end{eqnarray}
with the following convention $\Ad(\alpha_g(h),g') X = ghg^{-1}g'X{g'}^{-1}gh^{-1}g^{-1}$.

\subsection{The notion of compatible connections}
Before to examine the possibility to endow a trivial 2-bundle with a connective structure, we need a simple lemma.
\begin{lemma}
\label{lemma}
Let $P$ be a principal $G$-bundle ($G$ is a Lie group) over a manifold $M$ with transition functions $g^{ij}$. Let $f:G \to K$ be a group homomorphism. The right action of $K$ on itself defines a right action of $G$ on $K$ : $kf(g)$, $k\in K$ and $g\in G$. The associated bundle $P \times_{G,f} K = \{[(pg,f(g^{-1})k),g \in G]\}_{p\in P;k \in K}$ constitutes a principal $K$-bundle over $M$ with transition functions $f(g^{ij})$.
\end{lemma}
\begin{proof}
Let $\{U^i\}_i$ be a good open cover of $M$, and $\varphi^j : U^j \times G \to P_{|U^j}$ be the local trivializations of $P$. Let $\tilde \phi^j : U^j \times K \to (P \times_{G,f} K)_{|U^j}$ be the local trivializations of $P \times_{G,f} K$: $\tilde \phi^j(x,k) = [(\phi^j(x,g),f(g^{-1})k); g\in G] $. Since $\phi^j(x,g) = \phi^i(x,g^{ij}(x)g)$ (for $x \in U^i \cap U^j$), we have $\tilde \phi^j(x,k) = [\phi^i(x,g^{ij}(x)g),f(g^{-1})k);g\in G]$. By the variable change $\hat g = g^{ij}g$ we have 
$$\tilde \phi^j(x,k) = [(\phi^i(x,\hat g),f(\hat g^{-1})f(g^{ij}(x))k);\hat g \in G] = \tilde \phi^i(x,f(g^{ij})k)$$
\end{proof}

We want endow $P \times_{G,f} K$ with a connection which would be viewed as an image of a connection of $P$. The action of $G$ on $K$ being not necessary faithful, we require only a notion of compatibility between the two connections:
\begin{definition}[Compatible connections]
Let $HP$ and $H(P \times_{G,f} K)$ be connections (horizontal tangent spaces) of $P$ and $P \times_{G,f} K$. Let $j : P \to P \times_{G,f} K$ be the map defined by $\forall p \in P$, $j(p) = [(pg,f(g^{-1}));g\in G]$, i.e. $j(P) = P \times_{G,f} \{e_K\} \subset P \times_{G,f} K$. We say that the two connections are compatible if $j_* H_pP = H_{j(p)}(P \times_{G,f} K)$, where $j_*$ is the push-forward of $j$. Let $\omega \in \Omega^1(P,\mathfrak g)$ and $\tilde \omega \in \Omega^1(P \times_{G,f} K,\mathfrak k)$ be the associated connection 1-forms ($\ker \omega = HP$), we have then $j^* \tilde \omega = f^{Lie}(\omega) \in \Omega^1(P,\mathfrak k)$ where $f^{Lie}$ is Lie algebra homomorphism induced by $f$ and $j^*$ is the pull-back of $j$.
\end{definition}

\subsection{2-connection on a trivial 2-bundle}
We consider a principal $\mathcal G$-2-bundle $\mathcal P$ over an affine 2-space $\mathcal M$. In this section we suppose that $\mathcal P$ is trivial, in the sense where its 2-transition functions are trivial : $h^{ijk}(x) = e_H$. In that case, the $G$-transition functions satisfy the cocycle relation $g^{ij}(x)g^{jk}(x)=g^{ik}(x)$ ($\forall x \in U^i \cap U^j \cap U^k$) and define then a principal $G$-bundle $P$. We call it the object-bundle. The $H$-transition functions satisfy $h^{ij}(y,x) \alpha_{g^{ij}(x)}(h^{jk}(y,x)) = h^{ik}(y,x)$ ($\forall x \in U^i \cap U^j \cap U^k$). Let $q^{ij}(y,x) = (h^{ij}(y,x),g^{ij}(x)) \in \underline{H \rtimes G}_{(U^i \cap U^j)^2}$. We see that $q^{ij}$ satisfy the cocycle relation $q^{ij}(y,x)q^{jk}(y,x) = q^{ik}(y,x)$ ($\forall x,y \in U^i \cap U^j \cap U^k$). $q^{ij}$ can be then viewed as the transition functions of a principal $H \rtimes G$-bundle $Q$ over $M^2_\triangle = \bigcup_i (U^i \times U^i)_{/\mathcal R} \subset M^2_{/\mathcal R}$. We call it the arrow-bundle. We denote by $\pi_P$ and $\pi_Q$ the projections of $P$ and $Q$.\\

$\phi^i : \mathcal U^i \times \mathcal G \to \mathcal P_{|\mathcal U^i}$ being a functor, we have by definition
\begin{equation}
\begin{cases}
\id_{\phi^i(x,g)}  = \phi^i(\overleftarrow{xx},e_H,g) & \\
s(\phi^i(\overleftarrow{yx},h,g)) = \phi^i(x,g) & \\
t(\phi^i(\overleftarrow{yx},h,g)) = \phi^i(y,t(h)g) & 
\end{cases}
\end{equation}
We want formulate these relations in the langage of the bundle theory in the case where $\mathcal P$ is trivial.\\
Let $\Delta : \begin{array}{rcl} M & \to & M^2_\triangle \\ x & \mapsto & (x,x) \end{array}$ be the diagonal map. Let $\Delta^* Q = \{(x,q) \in M \times Q | \Delta(x)=\pi_Q(q) \}$ be the $H \rtimes G$-bundle over $M$ induced by $Q$ via $\Delta$. By construction the transition functions of $\Delta^*Q$ are $g_I^{ij}(x) = q^{ij}(\Delta(x)) = (e_H,g^{ij}(x))$. Clearly this is the transition functions of the widening of $P$, i.e. $P \times_G (H \rtimes G)$ (where we have considered $G \simeq \{e_H\} \rtimes G$ as a subgroup of $H \rtimes G$, the isomorphism between $G$ and $\{e_H\}\rtimes G$ constituting the homomorphism for the lemma \ref{lemma}). We denote $I = \Delta^* Q = P \times_G (H \rtimes G)$ and we call it the identity-bundle. Let $\Delta_* : \begin{array}{rcl} I & \to & Q \\ (x,q) & \mapsto & q \end{array}$ and  $\iota : P \to I$ be defined by $\forall p \in P$, $\iota(p) = [(pg,e_H,g^{-1});g \in G]$. We have then the following commutative diagram
$$ \begin{CD}
P @>{\iota}>> I @>{\Delta_*}>> Q \\
@VVV @VVV @VVV \\
M @= M @>{\Delta}>> M^2_{\triangle}
\end{CD} $$
We denote by $\Delta_{**} : TI \to TQ$ and by $\Delta^*_* : \Omega^* Q \to \Omega^* I$ the push-forward and the pull-back of $\Delta_*$, and by $\iota_* : TP \to TI$ and $\iota^* : \Omega^* I \to \Omega^* P$ the push-forward and the pull-back of $\iota$.\\
Let $\Pi_1 : \begin{array}{rcl} M^2_{\triangle} & \to & M \\ (y,x) & \mapsto & y \end{array}$. Let $\Pi_1^* P = \{(y,x,p)\in M^2_\triangle \times P | y=\pi_P(p) \}$ be the $G$-bundle over $M^2_\triangle$ induced by $P$ via $\Pi_1$. By construction the transition functions of $\Pi_1^* P$ are $g_T^{ij}(y,x) = g^{ij}(\Pi_1(y,x))=g^{ij}(y)$. Let $t:H \rtimes G \to  G$ be the homomorphism defined by $t(h,g)=t(h)g$. The bundle $Q \times_{H \rtimes G,t}G$ defined by $t$ and the lemma \ref{lemma} has the same transition functions $t(q^{ij}(y,x)) = t(h^{ij}(y,x))g^{ij}(x) = g^{ij}(y)$. We have then $T = \Pi_1^* P = Q \times_{H \rtimes G,t} G$ that we call it the target-bundle. Let $\Pi_{1*} : \begin{array}{rcl} T & \to & P \\ (y,x,p) & \mapsto & p \end{array}$ and $\tau : Q \to T$ be defined by $\forall q \in Q$, $\tau(q) = [(q(h,g),g^{-1}t(h^{-1}));h\in H, g \in G ]$. We have then the following commutative diagram
$$ \begin{CD}
P @<{\Pi_{1*}}<< T @<{\tau}<< Q \\
@VVV @VVV @VVV \\
M @<{\Pi_1}<< M^2_\triangle @= M^2_\triangle
\end{CD} $$
We denote by $\Pi_{1**} : TT \to TP$ and by $\Pi_{1*}^* : \Omega^* P \to \Omega^* T$ the push-forward and the pull-back of $\Pi_{1*}$, and by $\tau_* : TQ \to TT$ and $\tau^* : \Omega^* T \to \Omega^* Q$ the push-forward and the pull-back of $\tau$.\\
Let $\Pi_2 : \begin{array}{rcl} M^2_\triangle & \to & M \\ (y,x) & \mapsto & x \end{array}$. Let $\Pi_2^* = \{(y,x,p)\in M^2_\triangle \times P | x = \pi_P(p) \}$ be the $G$-bundle over $M^2_\triangle$ induced by $P$ via $\Pi_2$. By construction the transition functions of $\Pi_2^* P$ are $g_S^{ij}(y,x) = g^{ij}(x)$. Let $s:H \rtimes G$ be the homomorphism defined by $s(h,g)=g$. The bundle $Q \times_{H \rtimes G,s}G$ defined by $s$ and the lemma \ref{lemma} has the same transition functions $s(q^{ij}(y,x))=g^{ij}(x)$. We have then $S = \Pi_2^* P = Q \times_{H \rtimes G,s} G$ that we call the source-bundle. Let $\Pi_{2*} : \begin{array}{rcl} S & \to & P \\ (y,x,p) & \mapsto & p \end{array}$ and $\varsigma : Q \to S$ be defined by $\forall q \in Q$, $\varsigma(q) = [(q(h,g),g^{-1});h\in H, g \in G ]$. We have then the following commutative diagram
$$ \begin{CD}
P @<{\Pi_{2*}}<< S @<{\varsigma}<< Q \\
@VVV @VVV @VVV \\
M @<{\Pi_2}<< M^2_\Delta @= M^2_\Delta
\end{CD} $$
We denote by $\Pi_{2**} : TS \to TP$ and by $\Pi_{2*}^* : \Omega^* P \to \Omega^* S$ the push-forward and the pull-back of $\Pi_{2*}$, and by $\varsigma_* : TQ \to TS$ and $\varsigma^* : \Omega^* S \to \Omega^* Q$ the push-forward and the pull-back of $\varsigma$.\\ 
We denote by $\varphi^i_P$ and $\varphi^i_Q$ the transition functions of $P$ and $Q$ (they corresponds respectively to the object and the arrow parts of the functor $\phi^i$). The three previous commutative diagrams can be rewritten as follows:
\begin{equation}
\begin{cases}
\Delta_* \circ \iota \left(\varphi^i_P(x,g) \right) = \varphi^i_Q(x,x,e_H,g) & \\
\Pi_{2*} \circ \varsigma \left(\varphi^i_Q(y,x,h,g)\right) = \varphi^i_P(x,g) & \\
\Pi_{1*} \circ \tau \left(\varphi^i_Q(y,x,h,g)\right) = \varphi^i_P(y,t(h)g) &
\end{cases}
\end{equation}
This is the reformulation of the functor properties of $\phi^i$ in the fiber bundle language.\\

Since $P$ and $Q$ are principal bundles, they have canonical vertical tangent spaces : $T_pP \supset V_pP \simeq \mathfrak g$ ($\forall p \in P)$ and $T_qQ \supset V_qQ \simeq \mathfrak h \soplus \mathfrak g$ ($\forall q \in Q$). We define a connection of $\mathcal P$ as being two connections, one of $P$ and one of $Q$, compatible with the category structure of $\mathcal P$, and then compatible with the commutative diagrams linking $P$ and $Q$ via $I$, $T$ and $S$. 

\begin{definition}[2-connection on a trivial 2-bundle]
A 2-connection on a trivial 2-bundle $\mathcal P$, is the data of a connection $HP$ on $P$ and a connection $HQ$ on $Q$ such that the horizontal spaces satisfy
\begin{eqnarray}
\forall p \in P \qquad \Delta_{**} \iota_* H_pP & = & H_{\Delta_* \circ \iota(p)}Q \\
\forall q \in Q \qquad \Pi_{1**} \tau_* H_qQ & = & H_{\Pi_{1*} \circ \tau (q)} P \\
\forall q \in Q \qquad \Pi_{2**} \varsigma_* H_qQ & = & H_{\Pi_{2*} \circ \varsigma(q)} P
\end{eqnarray}
\end{definition}
This definition can be expressed in the terms of the connection 1-forms. Let $\omega_P \in \Omega^1(P,\mathfrak g)$ be the connection 1-form of $P$ ($\ker \omega_P = HP$) and $\omega_Q \in \Omega^1(Q,\mathfrak h \soplus \mathfrak g)$ be the connection 1-form of $Q$ ($\ker \omega_Q = HQ$). We have then :
\begin{eqnarray}
\iota^* \Delta^*_* \omega_Q & = & \omega_P \\
\varsigma^* \Pi^*_{2*} \omega_P & = & \pi^{\mathfrak g}(\omega_Q) \\
\tau^* \Pi_{1*}^* \omega_P & = & t^{Lie}_\bullet(\omega_Q)
\end{eqnarray}
with $t^{Lie}_\bullet = t^{Lie} \oplus \id_{\mathfrak g}$ defined on $\mathfrak h \soplus \mathfrak g$.\\

We consider now the local data of the connection. Let $\sigma^i_P \in \Gamma(U^i,P)$ be the trivializing local section of $P$ 
\begin{equation}
\forall x \in U^i \cap U^j, \quad \sigma^j_P(x) = \sigma^i_P(x) g^{ij}(x)
\end{equation}
and $\sigma^i_Q \in \Gamma(U^i \times U^i,Q)$ be the trivializing local section of $Q$ :
\begin{equation}
\forall x,y\in U^i, \quad \sigma_Q^j(y,x) = \sigma_Q^i(y,x) q^{ij}(y,x)
\end{equation}
By the properties of the local trivializations we have
\begin{equation}
\begin{cases}
\Delta_* \circ \iota \left(\sigma^i_P(x) \right) = \sigma^i_Q(x,x) & \\
\Pi_{2*} \circ \varsigma \left(\sigma^i_Q(y,x) \right) =  \sigma^i_P(x) & \\
\Pi_{1*} \circ \tau \left(\sigma^i_Q(y,x) \right) =  \sigma^i_P(y) &
\end{cases}
\end{equation}

We can then define a $G$-gauge potential $A^i = \sigma^{i*}_P \omega_P \in \Omega^1(U^i,\mathfrak g)$ and a $H \rtimes G$-gauge potential $\underline \eta^i = \sigma^{i*}_Q \omega_Q \in \Omega^1(U^i \times U^i_{/\mathcal R},\mathfrak h \soplus \mathfrak g)$. The relations between the connection 1-forms and between the trivializing local sections induce
\begin{eqnarray}
\Delta^* \underline \eta^i(x) & = & A^i(x) \\
\pi^{\mathfrak g}(\underline \eta^i(y,x)) & = & A^i(x) \\
t^{Lie}_\bullet(\underline \eta^i(y,x)) & = & A^i(y)
\end{eqnarray}
This induces that $\underline \eta^i(y,x) = A^i(x) + \eta^i(y,x)$ with $\eta^i \in \Omega^1(U^i \times U^i_{/\mathcal R}, \mathfrak h)$, such that $\eta^i(x,x) = 0$ and
\begin{equation}
t^{Lie}(\eta^i(y,x)) = A^i(y) - A^i(x) \in \Omega^1(U^i \times U^i_{/\mathcal R}, t^{Lie}(\mathfrak h))
\end{equation}

 By construction we have
\begin{equation}
\forall x \in U^i \cap U^j, \quad A^j(x) = g^{ij}(x)^{-1} A^i(x) g^{ij}(x) + g^{ij}(x)^{-1}dg^{ij}(x)
\end{equation}
\begin{equation}
\forall x,y \in U^i \cap U^j, \quad \underline \eta^j(y,x) = q^{ij}(y,x)^{-1} \underline \eta^i(y,x) q^{ij}(y,x) + q^{ij}(y,x)^{-1}\dd q^{ij}(y,x)
\end{equation}
where $\dd = d_x + d_y =\frac{\partial}{\partial x^\mu} dx^\mu + \frac{\partial}{\partial y^\nu} dy^\nu$ denotes the exterior differential of $M^2_\triangle$.
\begin{property}
$\forall x,y \in U^i \cap U^j$ we have
\begin{eqnarray}
\alpha_{g^{ij}(x)}\left(\eta^j(y,x) \right) & = & h^{ij}(y,x)^{-1} \eta^i(y,x) h^{ij}(y,x) \nonumber \\
& & \qquad + h^{ij}(y,x)^{-1}\dd h^{ij}(y,x) \nonumber \\
& & \qquad + h^{ij}(y,x)^{-1} \alpha^{Lie}_{A^i(x)}(h^{ij}(y,x))
\end{eqnarray}
\end{property}

\begin{proof}
We have 
\begin{eqnarray*}
q^{ij-1}\underline \eta^iq^{ij} & = & g^{ij-1}h^{ij-1}A^i h^{ij}g^{ij} + g^{ij-1}h^{ij-1}\eta^i h^{ij}g^{ij} \\
&  = & g^{ij-1}A^ig^{ij} + g^{ij-1}h^{ij-1}[A^i,h^{ij}]g^{ij} + g^{ij-1}h^{ij-1}\eta^i h^{ij}g^{ij}
\end{eqnarray*}
Moreover 
$$q^{ij-1}\dd q^{ij} = g^{ij-1}h^{ij-1}\dd h^{ij} g^{ij} = g^{ij-1}(h^{ij-1}\dd h^{ij})g^{ij}+g^{ij-1}dg^{ij}$$
We have then 
$$\underline \eta^j =g^{ij-1}A^i g^{ij} + g^{ij-1}dg^{ij} + g^{ij-1} \left(h^{ij-1}\eta^i h^{ij} + h^{ij-1}\dd h^{ij} + h^{ij-1} [A^i,h^{ij}] \right)g^{ij}$$
\end{proof}

Finally we can introduce the curvatures of the connections, $F^i = dA^i + A^i \wedge A^i \in \Omega^2(U^i,\mathfrak g)$ and $\underline F^i = \dd \underline \eta^i + \underline \eta^i \wedge \underline \eta^i \in \Omega^2(U^i\times U^i_{/\mathcal R},\mathfrak h \soplus \mathfrak g)$. It is easy to see that $\underline F^i$ can be decomposed as $\underline F^i = F^i + B^i$ with
\begin{equation}
B^i = \dd \eta^i + \eta^i \wedge \eta^i + \alpha^{Lie}_{A^i}(\eta^i) \in \Omega^2(U^i \times U^i_{/\mathcal R}, \mathfrak h)
\end{equation}
We can note that $F^i$ is equivariant : $\forall x,y \in U^i \cap U^j$ 
\begin{equation}
F^j = {g^{ij}}^{-1} F^i g^{ij} 
\end{equation}
but the curving satisfies
\begin{equation}
\alpha_{g^{ij}}(B^j) = {h^{ij}}^{-1} B^i h^{ij} + {h^{ij}}^{-1} \alpha^{Lie}_{F^i}(h^{ij})
\end{equation}
In the higher gauge theory litterature, $\underline F^i$ (or $F^i + t^{Lie}(B^i)$) is usually called the fake curvature and $B^i$ is usually called the curving. We can also introduce the true curvature (or 3-curvature) $H^i = \dd B^i + \alpha^{Lie}_{A^i}(B^i) = \dd \underline F^i + [A^i,\underline F^i] = - [\eta^i,\underline F^i] \in \Omega^3(U^i\times U^i_{/\mathcal R},\mathfrak h)$ which satisfies the generalized Bianchi identity $\dd H^i + \alpha^{Lie}_{A^i}(H^i) + [B^i,\underline F^i] = 0$.\\


If $\mathcal M$ is hyperbolic, there exists also $H \rtimes G$-bundles over $M^n_\triangle = \bigcup_i (U^i)^n_{/\mathcal R}$ ($n \geq 3$) denoted by $Q^{(n-1)}$ with the transition functions $q^{ij}_{n-1}(z,...,x) = q^{ij}(z,x)$. We consider the case of $Q^{(2)}$. The relations between $Q^{(2)}$ and $P$ are the same than between $Q$ and $P$. Moreover $Q^{(2)}$ is related to $Q$ by the partial diagonal maps: $\Delta_t : (y,x) \mapsto (y,y,x)$ and $\Delta_s:(y,x) \mapsto (y,x,x)$. This induces that $Q^{(2)}$ is endowed with a connection of gauge potential $\underline {\underline \eta}^i \in \Omega^1((U^i)^3_{/\mathcal R},\mathfrak h \soplus \mathfrak g)$ such that
\begin{eqnarray}
\Delta_0^* \underline {\underline \eta}^i(x) & = & A^i(x) \\
\Delta_s^* \underline {\underline \eta}^i(y,x) & = & \underline \eta^i(y,x) \\
\Delta_t^* \underline {\underline \eta}^i(y,x) & = & \underline \eta^i(y,x) \\
t^{Lie}_\bullet(\underline {\underline \eta}^i(z,y,x)) & = & A^i(z) \\
\pi^{\mathfrak g} (\underline {\underline \eta}^i(z,y,x)) & = & A^i(x)
\end{eqnarray}
with $\Delta_0: x \mapsto (x,x,x)$ the diagonal map. This induces that $\underline {\underline \eta}^i(z,y,x) = \underline \eta^i(z,y) + \underline \eta^i(y,x) - A^i(y) = \eta^i(z,y)+\eta^i(y,x)+A^i(x)$. The connection of $Q^{(2)}$ does not contain new information. This argument can be repeated for $Q^{(n>2)}$.

\subsection{2-connection : general case}
Let $\{P^i\}_i$ and $\{Q^i\}_i$ be the local principal $G$-bundles and $H \rtimes G$-bundles over $U^i$ and $U^i\times U^i_{/\mathcal R}$ defined by $P^i = \{\phi^i(x,g); x \in U^i,g\in G\}$ and $Q^i = \{ \phi^i(\overleftarrow{yx},h,g); (y,x)\in (U^i)^2_{/\mathcal R},h\in H,g\in G\}$. The 2-transition functions $h^{ijk}(x)$ constitute an obstruction to lift $\{P^i\}_i$ and $\{Q^i\}_i$ as globally defined principal bundles (because of the failure of the cocycle relations for $g^{ij}(x)$ and $h^{ij}(y,x)$). The construction followed in the previous section can nevertheless be reiterated over each 2-chart $\mathcal U^i$ but not globally. We have then local indentity-bundles $I^i$, local target-bundles $T^i$ and local source-bundles $S^i$. Nevertheless we need of a global bundle ensuring the global consistency of the connective structure. By definition of a Lie crossed module, $t(H)$ is a normal subgroup of $G$. We have then the following extension of groups:
$$ 1 \to H \xrightarrow{t} G \xrightarrow{\wp} G/t(H) \to 1 $$ 
Let $R$ be the principal $G/t(H)$-bundle over $M$ defined by the transition functions $\wp(g^{ij}(x))$ (since $t(h^{ijk}(x)) \in \ker \wp$, $\wp(g^{ij}(x))$ satisfies the cocycle relation $\wp(g^{ij}(x)g^{jk}(x)) = \wp(g^{ik}(x))$). Let $P^i/t(H)$ be the principal $G/t(H)$-bundle over $U^i$ induced by $\wp$ with $P^i$. We denote by $\vartheta^i : P^i \to P^i/t(H)$ the map defined by $\forall p \in P^i$, $\vartheta^i(p) = pt(H)$ (where the canonical right action of $G$ on $P^i$ is simply denoted by a right multiplication). Clearly, $P^i/t(H)$ and $R$ are diffeormophic over $U^i$: $P^i/t(H) \simeq R_{|U^i}$. Moreover we have the following commutative diagram over $U^i \cap U^j$ :
$$ \begin{CD}
P^j_{|U^i \cap U^j} @>{\phi^i \bar \phi^j}>> P^i_{|U^i \cap U^j} \\
@V{\vartheta^j}VV @VV{\vartheta^i}V \\
R_{|U^i \cap U^j} @>{\varphi^i_R \varphi^{j-1}_R}>> R_{|U^i \cap U^j}
\end{CD} $$
where $\varphi^i_R$ are the local trivializations of $R$.
\begin{definition}[2-connection on a 2-bundle]
A 2-connection on a 2-bundle $\mathcal P$, is the data of connections $HP^i$ on $P^i$, connections $HQ^i$ on $Q^i$ and a connection $HR$ on $R$ such that the horizontal spaces satisfy
\begin{eqnarray}
\forall p\in P^i \qquad \Delta_{**} \iota_* H_pP^i & = & H_{\Delta_* \circ \iota(p)} Q^i \\
\forall q\in Q^i \qquad \Pi_{2**} \varsigma_* H_q Q^i & = & H_{\Pi_{2*} \circ \varsigma(q)} P^i \\
\forall q\in Q^i \qquad \Pi_{1**} \tau_* H_q Q^i & = & H_{\Pi_{1*} \circ \tau(q)} P^i \\
\forall p\in P^i \qquad \qquad \vartheta^i_* H_pP^i & = & H_{\vartheta^i(p)} R
\end{eqnarray}
\end{definition}
We can explain the relations between the connections by using the connection 1-forms $\omega^i_P \in \Omega^1(U^i,\mathfrak g)$, $\omega^i_Q \in \Omega^1(U^i\times U^i_{/\mathcal R}, \mathfrak h \soplus \mathfrak g)$ and $\omega_R \in \Omega^1(R,\wp^{Lie}(\mathfrak g))$ ($\wp^{Lie} : \mathfrak g \to \mathfrak g/t^{Lie}(\mathfrak h)$ is Lie algebra homomorphism induced by $\wp$):
\begin{eqnarray}
\iota^* \Delta^*_* \omega_Q^i & = & \omega_P^i \\
\varsigma^* \Pi^*_{2*} \omega_P^i & = & \pi^{\mathfrak g}(\omega_Q^i) \\
\tau^* \Pi^*_{1*} \omega_P^i & = & t^{Lie}_\bullet(\omega_Q^i) \\
\vartheta^{i*} \omega_R & = & \wp^{Lie}(\omega^i_P)
\end{eqnarray}
Let $\sigma^i_P \in \Gamma(U^i,P^i)$ and $\sigma^i_Q \in \Gamma(U^i \times U^i_{/\mathcal R},Q^i)$ be the trivializing local sections : $\sigma^i_P(x) = \phi^i(x,e_G)$ and $\sigma^i_Q(y,x) = \phi^i(\overleftarrow{yx},e_H,e_G)$. We can define the $G$-gauge potential $A^i = \sigma^{i*}_P \omega_P^i \in \Omega^i(U^i,\mathfrak g)$ and the $H \rtimes G$-gauge potential $\underline \eta^i = \sigma^{i*}_Q \omega^i_Q \in \Omega^1(U^i \times U^i_{/\mathcal R}, \mathfrak h \soplus \mathfrak g)$. By the same arguments that for the case of the trivial 2-bundles, we have $\underline \eta^i (y,x) = A^i(x) + \eta^i(y,x)$ with $\eta^i \in \Omega^1(U^i \times U^i_{/\mathcal R},\mathfrak h)$ and $t^{Lie}(\eta^i(y,x)) = A^i(y) - A^i(x)$. \\
Consider the 1-form on $U^i \cap U^j$ defined by $\sigma^{j*}_P \omega^j_P - \sigma^{j*}_P \bar \phi^{i*} \phi^{i*} \omega^i_P$. Since $\phi^i \bar \phi^i \sigma^j_P(x) = \phi^i \bar \phi^i \phi^j(x,e_G) = \phi^i(x,g^{ij}(x)) = R(g^{ij}(x)) \sigma^i_P(x)$ we have
\begin{equation}
\sigma^{j*}_P \omega^j_P - \sigma^{j*} \bar \phi^{i*} \phi^{i*} \omega^i_P = A^j(x) - g^{ij}(x)^{-1} A^i(x) g^{ij}(x) - g^{ij}(x)^{-1}dg^{ij}(x)
\end{equation}
But we have also
\begin{equation}
\wp^{Lie}(\sigma^{j*}_P \omega^j_P - \sigma^{j*} \bar \phi^{i*} \phi^{i*} \omega^i_P) = \sigma^{j*}_P \vartheta^{j*} \omega_R - \sigma^{j*}_P \bar \phi^{i*} \phi^{i*} \vartheta^{i*} \omega_R
\end{equation}
But $\vartheta^i \phi^i \bar \phi^i = \varphi^i_R \varphi^{i-1}_R \vartheta^i = \vartheta^i$ and $\vartheta^i \sigma^j_P(x) = \vartheta^i\sigma^i_P(x) \wp(g^{ij}(x))$. We have then
\begin{eqnarray}
& & \wp^{Lie}(\sigma^{j*}_P \omega^j_P - \sigma^{j*}_P \bar \phi^{i*} \phi^{i*} \omega^i_P) \nonumber \\
& & \quad  = a^j(x) - \wp(g^{ij}(x))^{-1} a^i(x) \wp(g^{ij}(x)) - \wp(g^{ij}(x))^{-1}d\wp(g^{ij}(x)) \\
& & \quad =  0
\end{eqnarray}
where $a^i = \sigma^{i*}_P \vartheta^{i*} \omega_R$ is the gauge potential of $R$ associated with the section $\vartheta^i \sigma^i_P(x) \in \Gamma(U^i,R)$. We conclude then that $A^j(x) - g^{ij}(x)^{-1} A^i(x) g^{ij}(x) - g^{ij}(x)^{-1}dg^{ij}(x) \in \ker \wp^{Lie} = t^{Lie}(\mathfrak h)$.
\begin{definition}[potential-transformation]
We call the potential-transformation of the 2-connection, the 1-form $\eta^{ij} \in \Omega^1(U^i \cap U^j,\mathfrak h)$ such that
\begin{equation}
A^j(x) = g^{ij}(x)^{-1}A^i(x)g^{ij}(x) + g^{ij}(x)^{-1}dg^{ij}(x) + t^{Lie}(\eta^{ij}(x))
\end{equation}
\end{definition} 
The gluing relation of the $P^i$-curvatures is not equivariant:
\begin{equation}
dA^j + A^j \wedge A^j = {g^{ij}}^{-1} (dA^i + A^i \wedge A^i) g^{ij} + t^{Lie}(d\eta^{ij} + \alpha^{Lie}_{A^j}(\eta^{ij}) - \eta^{ij} \wedge \eta^{ij})
\end{equation}
but $\wp^{Lie}(dA^i + A^i \wedge A^i)$ is equivariant (with $\wp(g^{ij})$). Let $B^i_{sph} \in \Omega^2(U^i,\mathfrak h)$ be a 2-form such that 
\begin{equation}
B^j_{sph}(x) - \alpha_{g^{ij}(x)^{-1}}(B^i_{sph}(x)) = d\eta^{ij}(x) + \alpha^{Lie}_{A^j(x)}(\eta^{ij}(x)) - \eta^{ij}(x) \wedge \eta^{ij}(x)
\end{equation}
Then $F^i = dA^i + A^i \wedge A^i - t^{Lie}(B^i_{sph}) \in \Omega^2(U^i,\mathfrak g)$ is equivariant and belongs to the equivalence class of the 2-forms of $P^i$ compatible with the curvature of $R$ ($\wp(F^i) = \wp(dA^i + A^i \wedge A^i) = da^i + a^i \wedge a^i$). We consider then $F^i$ as being the fake curvature of $\{P^i\}$. We call $B^i_{sph}$ the spherical part of the curving.
\begin{property}
The gluing relation of the $H \rtimes G$-gauge potential $\underline \eta^i \in \Omega^1(U^i \times U^i_{/\mathcal R},\mathfrak h \soplus \mathfrak g)$ is
\begin{eqnarray}
\underline \eta^j(y,x) & = & q^{ij}(y,x)^{-1} \underline \eta^i(y,x) q^{ij}(y,x) + q^{ij}(y,x)^{-1} \dd q^{ij}(y,x) \nonumber \\
& & \qquad + t^{Lie}(\eta^{ij}(x)) + \eta^{ij}(y) - \eta^{ij}(x)
\end{eqnarray}
\end{property}
\begin{proof}
Let $\nu^{ij} \in \Omega^2((U^i \cap U^j)^2_{/\mathcal R},\mathfrak h \soplus \mathfrak g)$ be $\nu^{ij} = \underline \eta^j - q^{ij-1} \underline \eta^i q^{ij} - q^{ij-1} \dd q^{ij}$. By using the properties of $\underline \eta^i$ under the actions of $\Delta^*$, $t^{Lie}$ and $\pi^{\mathfrak g}$, and the gluing relation of $A^i$ we find
\begin{eqnarray}
\nu^{ij}(x,x) & = & t^{Lie}(\eta^{ij}(x)) \\
t^{Lie}_\bullet(\nu^{ij}(y,x)) & = & t^{Lie}(\eta^{ij}(y)) \\
\pi^{\mathfrak g}(\nu^{ij}(y,x)) & = & t^{Lie}(\eta^{ij}(x))
\end{eqnarray}
This induces that $\nu^{ij}(y,x) = t^{Lie}(\eta^{ij}(x)) + \eta^{ij}(y)- \eta^{ij}(x)$ (modulo an ignored element of $\ker t^{Lie}$ without significance). 
\end{proof}

By following the same arguments as for a trivial 2-bundle, we have $\underline \eta^i(y,x) = A^i(x) + \eta^i(y,x)$, with $\eta^i \in \Omega^1(U^i\times U^i_{/\mathcal R},\mathfrak h)$ satisfying the gluing relation: $\forall (x,y) \in (U^i\cap U^j)_{/\mathcal R}$
\begin{eqnarray}
\alpha_{g^{ij}(x)}(\eta^j(y,x)) & = & h^{ij}(y,x)^{-1} \eta^i(y,x) h^{ij}(y,x) \nonumber \\
& & \qquad + h^{ij}(y,x)^{-1} \dd h^{ij}(y,x) \nonumber \\
& & \qquad + h^{ij}(y,x)^{-1} \alpha^{Lie}_{A^i(x)}(h^{ij}(y,x)) \nonumber \\
& & \qquad + \alpha_{g^{ij}(x)}(\eta^{ij}(y)-\eta^{ij}(x))
\end{eqnarray}
Let $\underline F^i = \dd \underline \eta^i + \underline \eta^i \wedge \underline \eta^i \in \Omega^2(U^i \times U^i_{/\mathcal R},\mathfrak h \soplus \mathfrak g)$ be the fake curvature. $\underline F^i$ can be decomposed as $\underline F^i = F^i + t^{Lie}(B^i_{sph}) + B^i_{ns}$ with $B^i_{ns} = \dd \eta^i + \eta^i \wedge \eta^i + \alpha^{Lie}_{A^i}(\eta^i) \in \Omega^2(U^i \times U^i_{/\mathcal R},\mathfrak h)$ called the nonspherical part of the curving. $B^i_{sph}(x)+B^i_{ns}(y,x)$ forms the total curving. The gluing relation for the nonspherical part of the curving is
\begin{eqnarray}
B^j_{ns}(y,x) & = & \alpha_{g^{ij}(x)^{-1}} \left(h^{ij}(y,x)^{-1} B^i_{ns}(y,x) h^{ij}(y,x) + h^{ij}(y,x)^{-1} \alpha^{Lie}_{F^i(x)}(h^{ij}(y,x)) \right) \nonumber \\
& & \qquad + d\eta^{ij}(y) + \alpha^{Lie}_{A^j(x)}(\eta^{ij}(y)) - \eta^{ij}(y) \wedge \eta^{ij}(y) \nonumber \\
& & \qquad - d\eta^{ij}(x) - \alpha^{Lie}_{A^j(x)}(\eta^{ij}(x)) + \eta^{ij}(x) \wedge \eta^{ij}(x) \nonumber \\
& & \qquad + [\eta^i(y,x),\eta^{ij}(y)]
\end{eqnarray}
and for the total curving
\begin{eqnarray}
B^j(y,x) & = & \alpha_{g^{ij}(x)^{-1}} \left(h^{ij}(y,x)^{-1} B^i(y,x) h^{ij}(y,x) \right) \nonumber \\
& & \qquad + \alpha_{g^{ij}(x)^{-1}} \left(h^{ij}(y,x)^{-1} \alpha^{Lie}_{F^i(x)+B^i_{sph}(x)}(h^{ij}(y,x) \right) \nonumber \\
& & \qquad + d\eta^{ij}(y) + \alpha^{Lie}_{A^j(x)}(\eta^{ij}(y)) - \eta^{ij}(y) \wedge \eta^{ij}(y) \nonumber \\
& & \qquad + [\eta^i(y,x),\eta^{ij}(y)]
\end{eqnarray}

\begin{property}
The gluing relation of the potential-transformation is $\forall x \in U^i \cap U^j \cap U^k$
\begin{equation}
\alpha_{g^{ij}}(\eta^{ij})+\alpha_{g^{ij}g^{jk}}(\eta^{jk})-{h^{ijk}}^{-1} \alpha_{g^{ik}}(\eta^{ik})h^{ijk} = {h^{ijk}}^{-1}dh^{ijk}+{h^{ijk}}^{-1} \alpha^{Lie}_{A^i}(h^{ijk})
\end{equation}
\end{property}

\begin{proof}
\begin{eqnarray*}
& & g^{ik}t^{Lie}(\eta^{ik})g^{ik-1} \nonumber \\
& = & g^{ik}A^k g^{ik-1} - A^i - dg^{ik}g^{ik-1} \\
& = & g^{ik} \left(g^{jk-1}A^jg^{jk}+g^{jk-1}dg^{jk}+t^{Lie}(\eta^{jk})\right)g^{ik-1}- A^i - dg^{ik}g^{ik-1} \\ 
& = & g^{ik} \left(g^{jk-1} \left(g^{ij-1}A^i g^{ij}+g^{ij-1}dg^{ij}+t^{Lie}(\eta^{ij})\right)g^{jk}+g^{jk-1}dg^{jk} \right. \nonumber \\
& & \qquad \left. +t^{Lie}(\eta^{jk})\right)g^{ik-1}- A^i - dg^{ik}g^{ik-1}
\end{eqnarray*}
After some algebra we find
\begin{eqnarray*}
& & g^{ij}g^{jk}g^{ik-1}g^{ik} t^{Lie}(\eta^{ik}) g^{ik-1}g^{ik}g^{jk-1}g^{ij-1} \nonumber \\
& & = A^i - g^{ij}g^{jk}g^{ik-1}A^ig^{ik}g^{jk-1}g^{ij-1} + d(g^{ij}g^{jk}g^{ik-1})g^{ik}g^{jk-1}g^{ij-1} \nonumber \\
& & \quad + g^{ij}t^{Lie}(\eta^{ij})g^{ij-1}+g^{ij}g^{jk}t^{Lie}(\eta^{jk})g^{jk-1}g^{ij-1}
\end{eqnarray*}
Finally by using the relation $g^{ik}g^{jk-1}g^{ij-1} = t(h^{ijk})$  we prove the property modulo an ignored element of $\ker t^{Lie}$ without significance.
\end{proof}

\begin{table}
\caption{\label{localdata} Local data of a 2-bundle with a 2-connection}
\begin{center}
\begin{tabular}{|c|ccc|ccc|}
\hline
    &        & $M$    &        &        &$M^2_{\triangle}$ & \\
    & 0-form & 1-form & 2-form & 0-form & 1-form & 2-form  \\
\hline
    &        & $A^i$  & $F^i$  &         &       &\\
$G$ & $g^{ij}$&       &        &         &       &\\
    &        &        &        &         &       &\\
\hline
    & $k^i$  &        &$B^i_{sph}$&       & $\eta^i$&$B^i_{ns}$ \\
$H$ &        &$\eta^{ij}$&        & $h^{ij}$&        &\\
    &$h^{ijk}$&        &        &        &        & \\
\hline
\end{tabular}
\end{center}
\end{table}

Table \ref{localdata} summarizes the different data defining a 2-connection on a 2-bundle. In the other categorical bundle constructions \cite{brylinski,murray2,mackaay,kalkkinen2,aschieri2,baez1,baez2,baez3,wockel,soncini} the second column of the table \ref{localdata} (concerning the local data over $M^2_\triangle$) is absent. For the categorical bundles over pathspaces \cite{cattaneo,chatterjee,chatterjee2} the curving $B$ is not a 2-form on the manifold $M$, nor on $M^2_\triangle$ but on the total space of an usual principal bundle over $M$. It is a connective structure different from the one presented in this paper, but due to its link with the arrows of the pathspace category, we can consider it as equivalent to our nonspherical part of the curving but without $H$-gauge potential $\eta^i$. For non-abelian gerbes with connection \cite{laurent,breen}, similar data to the second column of the table \ref{localdata} can be found, but in this context there are defined as forms of the same manifold $M$ because in this construction the non-trivial categorical aspects are in the ``fibers'' (the gerbes) and not in the base space (as in our construction).\\

\begin{example}
Let $M=\{z \in \mathbb C, |z|\leq r\}$ with $r>0$ a constant, and the matrix $H(z) = \left( {0 \atop z} {{\bar z} \atop {-2\imath r}} \right)$. For $z\in \mathring{M}$ ($|z|<r$), $H(z)$ has two different eigenvalues $\lambda_\pm(z) = \imath(-r\pm \sqrt{r^2 -|z|^2})$, but for $z \in \partial M$ ($|z|=r$), $H(z)$ has only one eigenvalue $\lambda_0=-\imath r$. The generalized eigenvectors associated with an eigenvalue $\lambda_a(z)$ are solutions of $(H(z)-\lambda_a(z))^n \phi_a(z)=0$ for $n\geq 1$. For $z \in \mathring M$, the two linearly independent generalized eigenvectors are usual eigenvectors: $H(z) \phi_\pm(z) = \lambda_\pm(z) \phi_\pm(z)$, but for $z \in \partial M$, $H(z)$ is not diagonalizable and its two linearly independent generalized eigenvectors are $\phi_0(z) = \frac{1}{\sqrt 2} \left(\imath\atop {z/r} \right)$ with $H(z)\phi_0(z) = \lambda_0(z) \phi_0(z)$, and $\phi_0(-z)$ with $H(z) \phi_0(-z) = \lambda_0 \phi_0(-z) -2 \lambda_0 \phi_0(z)$ (i.e. $(H(z)-\lambda_0(z))^2\phi_0(-z)=0$). We can note the collapsus between the two eigenvectors $\lim_{|z| \to r} \phi_\pm(z) = \phi_0(re^{\imath \arg z})$. Since the eigenvectors are defined up to non-zero factor, we consider the local gauge changes : $\tilde \phi_a(z) = g_a(z) \phi_a(z)$ with $g_a \in \mathbb C^*$ (with $\lim_{|z|\to r} g_\pm(z)=g_0(re^{\imath \arg z})$). But in order to ensure the concistency of the equation $H(z) \phi_0(-z) = \lambda_0 \phi_0(-z) -2 \lambda_0 \phi_0(z)$, it is necessary to consider the gauge change redefinitions $h(-z,z)$ such that $h(-z,z)g_0(z) = g_0(-z)$ (i.e. $H(z) g_0(-z) \phi_0(-z) = \lambda_0 g_0(-z) \phi_0(-z) -2 \lambda_0 h(-z,z)g_0(z) \phi_0(z)$). The generalized eigenvectors of $H(z)$ define a 2-bundle. Its base 2-space $\mathcal M$ is defined by $\Obj(\mathcal M) = M$ and $z\mathcal R z'$ if $z=z'$ or if $z=-z$ with $|z|=r$ (the arrows between $z$ and $-z$ being associated with the fact that $H(z)$ links $\phi_0(-z)$ to $\phi_0(z)$). The affine 2-space $\mathcal M$ is euclidean on its boundary ($\partial M$) and spherical on its interior ($\mathring M$). The Lie crossed module $(G,H,t,\alpha)$ is defined by $G=\mathbb C^* \times \mathbb C^*$, $H=\mathbb C^*$, $t$ is the diagonal map $t(h) = \left( {h \atop 0} {0 \atop h}\right)$, and $\alpha$ is trivial ($\alpha_g = \id_H$, $\forall g \in G$). The gauge changes $g(z)=\left({{g_+(z)} \atop 0} {0 \atop {g_-(z)}} \right)$ are elements of $G$, and the gauge change redefinition for $z \in \partial M$ are arrows $(h(-z,z),g(z))$ with target $t(h(-z,z))\left({{g_0(z)} \atop 0} {0 \atop {g_0(z)}} \right)  = \left({{g_0(-z)} \atop 0} {0 \atop {g_0(-z)}} \right)$. We endow the 2-bundle with a 2-connection defined by $A(z) = \left({{\frac{\langle \chi|d\phi_+(z) \rangle}{\langle \chi|\phi_+(z)\rangle}} \atop 0} {0 \atop {\frac{\langle \chi|d\phi_-(z) \rangle}{\langle \chi|\phi_-(z)\rangle}}} \right) \in \Omega^1(M,\mathfrak g)$, with $\chi = \frac{1}{\sqrt 2}\left(1 \atop 1 \right)$ ($A(z)$ measures the local variations of the eigenvectors with respect to the Schr\"odinger cat state $\chi$); and by $\eta(-z,z) = \frac{\langle \chi|d\phi_0(-z)\rangle}{\langle \chi|\phi_0(-z)\rangle} - \frac{\langle \chi|d\phi_0(z)\rangle}{\langle \chi|\phi_0(z)\rangle} \in \Omega^1(\partial M,\mathfrak h)$ ($\partial M^2_{/\mathcal R} \simeq \partial M$). We have then $t^{Lie}(\eta(-z,z)) = A(-z)-A(z)$ for $z \in \partial M$. By gauge changes we have, $\tilde A(z) = A(z) + g(z)^{-1}dg(z)$ and $\tilde \eta(-z,z) = \eta(-z,z) + h(-z,z)^{-1} dh(-z,z)$. $G/t(H) \simeq \mathbb C^*$ and $a(z) = \wp^{Lie}(A(z)) = A_+(z) - A_-(z)$ ($a(z) = 0$ for $z \in \mathring M$). This small example is interesting because it exhibits a trivial 2-bundle (a single 2-chart is sufficient to cover the base 2-space) with highly non-trival 2-connection ($\eta \not=0$ and $a\not=0$).
\end{example}

\section{Horizontal lifts}
\subsection{Pseudosurfaces}
The scheme of the construction of a 2-bundle over an affine 2-space shows that the natural objects which can be lift in the bundle are not the surfaces but geometric entities related to the categorical structure.
\begin{definition}[Pseudosurfaces]
Let $\mathcal M$ be an affine 2-space. A pseudosurface is a smooth map $\gamma : [0,1] \to \Morph(\mathcal M)$ such that $\gamma(u)$ is constant near $u=0$ and near $u=1$.
\end{definition}
By definition, all pseudosurface on a spherical affine 2-space is reduced to a path-identity in $M$ ($u \mapsto \id_{s(\gamma(u))}$).\\
The path on $M = \Obj(\mathcal M)$ defined by $u \mapsto  s(\gamma(u))$ is called the source-boundary of the pseudosurface $\gamma$, and the path $u \mapsto t(\gamma(u))$ is called the target-boundary. Let $\{[0,1] \ni u \mapsto x_i(u) \in M \}_{i=1,...,n}$ be the minimal set of smooth paths such that $\gamma(u) = \overleftarrow{x_n(u)...x_1(u)}$. $\mathrm{Skel}_\gamma(u) = (x_n(u),...,x_1(u))$ is called the skeleton of the pseudosurface. We note that a skeleton can have junctions in the case where $x_i(u) = x_{i+1}(u)$ for $u \leq u_*$ (for example). A pseudosurface which has a skeleton reduced to its boundary is said elementary. By definition, all pseudosurface on an euclidean affine 2-space is elementary. It is interesting to point out some special cases of pseudosurfaces:
\begin{itemize}
\item A pseudosurface is said impervious if its boundary is closed : $\gamma(0) = \id_{s(\gamma(0))}$ and $\gamma(1) = \id_{s(\gamma(1))}$. An impervious pseudosurface is well delimited.
\item A pseudosurface is said cyclic if it is impervious and if $\gamma(0)=\gamma(1)$.
\item A pseudosurface is said pinched if $\forall u \in [0,1]$, $s(\gamma(u)) = s(\gamma(0))$ or/and $t(\gamma(u)) = t(\gamma(0))$.
\end{itemize}
\begin{definition}[Composition laws of pseudosurfaces]
Let $\gamma_1$ and $\gamma_2$ be two pseudosurfaces such that (for two parametrizations $u \mapsto \gamma_1(u)$ and $u \mapsto \gamma_2(u)$) $\mathrm{Skel}_{\gamma_2}(1) = \mathrm{Skel}_{\gamma_1}(0)$. The horizontal composition of these two pseudosurfaces is the pseudosurface $\gamma_1 \ast \gamma_2$ defined by the skeleton
$$\forall u \in [0,1], \quad  \mathrm{Skel}_{\gamma_1 \ast \gamma_2}(u) = \begin{cases} \mathrm{Skel}_{\gamma_2}(2u) & \text{if } u\in [0,\frac{1}{2}] \\ \mathrm{Skel}_{\gamma_1}(2u-1) & \text{if } u\in [\frac{1}{2},1] \end{cases} $$
Let $\gamma_1$ and $\gamma_2$ be two pseudosurfaces such that $\forall u$, $s(\gamma_1(u)) = t(\gamma_2(u))$. The vertical composition of the two pseudosurfaces is the pseudosurface $(\gamma_1 \circ \gamma_2)(u) = \gamma_1(u) \circ \gamma_2(u)$ (in the r.h.s. $\circ$ denotes the arrows composition of $\Morph(\mathcal M)$).
\end{definition}
The pseudosurfaces define a category $\mathcal{PS}(\mathcal M)$ with the smooth paths of $M$ as objects, the pseudosurfaces as arrows and the vertical composition as arrows composition. 

\subsection{Horizontal lifts of pseudosurfaces included in a single 2-chart}
\begin{definition}[Horizontal lifts of an elementary pseudosurface]
Let $\mathcal P$ be a 2-bundle over a affine 2-space $\mathcal M$ endowed with a 2-connection. Let $\gamma:[0,1] \to \Morph(\mathcal M)$ be an elementary pseudosurface. A map $\tilde \gamma:[0,1] \to \Morph(\mathcal P)$ is said to be a horizontal lift of $\gamma$ if $\forall u \in [0,1]$, $\pi(\tilde \gamma(u)) = \gamma(u)$ and if $\gamma(u) \in \Morph(\mathcal U^i)$ we have $X_{\tilde \gamma}^i(u) \in H_{\tilde \gamma(u)}Q^i$, $X_{s(\tilde \gamma)}^i(u) \in H_{s(\tilde \gamma(u))}P^i$ and $X_{t(\tilde \gamma)}^i(u) \in H_{t(\tilde \gamma(u))}P^i$ where $X_{\tilde \gamma}^i \in TQ^i$ is the tangent vector of $\tilde \gamma$ viewed as a path in $Q^i$ and $X_{s(\tilde \gamma)}^i, X_{t(\tilde \gamma)}^i \in TP^i$ are the tangent vectors of $s(\tilde \gamma)$ and $t(\tilde \gamma)$ viewed as paths in $P^i$.
\end{definition}

\begin{theorem}
Let $\sigma^i \in \Funct(\mathcal U^i,\mathcal P)$ be the trivializing local section : $\sigma^i(x) = \phi^i(x,e_G) = \sigma^i_P(x) $ and $\sigma^i(\overleftarrow{yx}) = \phi^i(\overleftarrow{yx},e_H,e_G) = \sigma^i_Q(y,x)$. Let $\gamma$ be an elementary pseudosurface completely included in $\mathcal U^i$ ($\forall u \in [0,1]$, $\gamma(u) \in \varphi(U^i \times U^i_{/\mathcal R})$). The horizontal lift of $\gamma$ passing through $\sigma^i(\gamma(0))$ is
\begin{equation}
\tilde \gamma^i(u) = \phi^i(\gamma(u),\PP_\gamma e^{-\int_{(y(0),x(0))}^{(y(u),x(u))} \underline \eta^i(y,x)})
\end{equation}
$\PP_\gamma e^{-\int_{(y(0),x(0))}^{(y(u),x(u))} \underline \eta^i(y,x)} \in H \rtimes G$ is path-ordering exponential along the path $u \mapsto (t(\gamma(u)),s(\gamma(u))) \in U^i \times U^i_{/ \mathcal R}$ (in order to simplify the notation we have denoted $s(\gamma(u))$ by $x(u)$ and $t(\gamma(u))$ by $y(u)$). 
\end{theorem}
By definition of the path-ordering exponential, $\PP_\gamma e^{-\int_{(y(0),x(0))}^{(y(u),x(u))} \underline \eta^i(y,x)}$ is solution of
\begin{equation}
\frac{d \PP_\gamma e^{-\int_{(y(0),x(0))}^{(y(u),x(u))} \underline \eta^i(y,x)}}{du} = - \underline \eta^i(y(u),x(u)) \PP_\gamma e^{-\int_{(y(0),x(0))}^{(y(u),x(u))} \underline \eta^i(y,x)}
\end{equation}
\begin{proof}
By applying the horizontal lift formula (see \cite{nakahara}) in the local $H \rtimes G$-bundle $Q^i$ endowed with the connection $\omega_Q$ we have directly $\pi(\tilde \gamma(u))= \gamma(u)$ and $X^i_{\tilde \gamma}(u) \in H_{\tilde \gamma(u)} Q^i$. By construction of the source and of the target bundles we have $X^i_{s(\tilde \gamma)}(u) = \Pi_{2**} \varsigma_* X^i_{\tilde \gamma}(u)$ and $X^i_{t(\tilde \gamma)}(u) = \Pi_{1**} \tau_* X^i_{\tilde \gamma}(u)$. It follows that $\omega_P^i(X^i_{s(\tilde \gamma)}(u)) = \omega_P^i(\Pi_{2**} \varsigma_* X^i_{\tilde \gamma}(u)) = \varsigma^* \Pi_{2*}^* \omega_P^i(X^i_{\tilde \gamma}(u)) = \pi^{\mathfrak g} (\omega_Q^i(X^i_{\tilde \gamma}(u))) = 0$ since $X^i_{\tilde \gamma}(u) \in  H_{\tilde \gamma(u)} Q^i = \ker \omega_Q^i$. In the same manner $\omega_P^i(X^i_{t(\tilde \gamma)}(u)) = t^{Lie}_\bullet (\omega_Q^i(X^i_{\tilde \gamma}(u))) = 0$. We have then $X^i_{s(\tilde \gamma)}(u), X^i_{t(\tilde \gamma)}(u) \in \ker \omega_P^i = HP^i$.
\end{proof}

The horizontal lift of $\gamma$ passing through $q \in \Morph(\mathcal P)$ with $\pi(q) = \gamma(0)$ is then $\tilde \gamma^i_q(u) = R(h,g) \tilde \gamma^i(u)$ where $(h,g) \in H \rtimes G$ is such that $R(h,g) \sigma^i(\gamma(0)) = q$. 

$\PP_\gamma e^{-\int_{(y(0),x(0))}^{(y(u),x(u))} \underline \eta^i(y,x)}$ is an element of $H \rtimes G$ represented in the universal enveloping algebra of $\mathfrak h \soplus \mathfrak g$. It is more interesting to have an expression of the horizontal lift as a couple $(h(u),g(u))$ with $h(u) \in H$ and $g(u) \in G$. A such expression is simple in the case where $H$ is the center of $G$ ($1 \to H \xrightarrow{t} G \to G/H \to 1$ is then a central extension of groups, $t$ is just the canonical injection of $H$ in $G$ and $\alpha$ is just the conjugation $\alpha_g(h) = ghg^{-1}=h$).
\begin{property}
Let $1 \to H \xrightarrow{t} G \to G/H \to 1$ be a central extension of groups. Let $\gamma$ be an elementary pseudosurface completely included in $\mathcal U^i$. The group element of the horizontal lift of $\gamma$ is then
\begin{equation}
\label{hl1}
\PP_\gamma e^{-\int_{(y(0),x(0))}^{(y(u),x(u))} \underline \eta^i(y,x)} = \left(e^{- \int_{(y(0),x(0))}^{(y(u),x(u))} \eta^i(y,x)}, \PP_{s(\gamma)}e^{- \int_{x(0)}^{x(u)} A^i(x)} \right) \in H \rtimes G
\end{equation}
We suppose that $\forall u \in [0,1]$, $x(1)$ and $y(u)$ are linkable, and $x(u)$ and $y(0)$ are linkable. Let $\mathcal C_\gamma^2$ be the closed path in $U^i \times U^i$ defined by $[0,1] \ni u \mapsto (y(u),x(u))$ for its first part, $[0,1] \ni u \mapsto (y(1-u),x(1))$ for its second part, and $[0,1] \ni u \mapsto (y(0),x(1-u))$ for its last part, and let $\mathcal S_\gamma^2$ be a surface in $U^i \times U^i$ having $\mathcal C^2_\gamma$ as boundary ($\partial \mathcal S_\gamma^2 = \mathcal C^2_\gamma$). We call $\mathcal S_\gamma^2$ a surface of the second kind supported by the pseudosurface $\gamma$. We have then
\begin{eqnarray}
& & \PP_\gamma e^{-\int_{(y(0),x(0))}^{(y(1),x(1))} \underline \eta^i(y,x)} \nonumber \\
\label{hl2} & & = \left(e^{-\iint_{\mathcal S_\gamma^2} B^i_{ns}(y,x)} e^{-\int_{s(\gamma)} \eta^i_s(y(0),x)} e^{- \int_{t(\gamma)} \eta^i_t(y,x(1))}, \PP_{s(\gamma)}e^{-\int_{s(\gamma)} A^i(x)} \right)
\end{eqnarray}
Moreover we suppose that $\gamma$ is impervious. Let $\mathcal C_\gamma^1$ be the closed path in $U^i$ defined by $[0,1] \ni u \mapsto y(u)$ ($t(\gamma)$) for its first part and $[0,1] \ni u \mapsto x(1-u)$ ($s(\gamma)^{-1}$) for its second part, and let $\mathcal S_\gamma^1$ be a surface in $U^i$ having $\mathcal C^1_\gamma$ as boundary ($\partial \mathcal S_\gamma^1=\mathcal C^1_\gamma$). We call $\mathcal S_\gamma^1$ a surface of the first kind supported by the pseudosurface $\gamma$. We have then
\begin{eqnarray}
& & \PP_\gamma e^{-\int_{(y(0),x(0))}^{(y(1),x(1))} \underline \eta^i(y,x)} \nonumber \\
\label{hl3} & & = \left(e^{-\iint_{\mathcal S_\gamma^2} B^i_{ns}(y,x)} e^{- \iint_{\mathcal S_\gamma^1} B^i_{sph}(x)}, \PP_{s(\gamma)}e^{-\int_{s(\gamma)} A^i(x)} \right)
\end{eqnarray}
\end{property}
\begin{proof}
Equation (\ref{hl1}) follows from the decomposition $\underline \eta^i(y,x) = \eta^i(y,x) + A^i(x)$ with $A^i(x)\in \mathfrak g$ and $\eta^i(y,x) \in \mathfrak h$, where $\PP_{s(\gamma)}e^{- \int_{x(0)}^{x(u)} A^i(x)} \in G$ is the path-ordering exponential along the path $u \mapsto s(\gamma(u))=x(u) \in U^i$. $e^{- \int_{(y(0),x(0))}^{(y(u),x(u))} \eta^i(y,x)} \in H$ is written without the path-ordering since $H$ is abelian if it is the center of $G$ (the expression is then an usual exponential).\\
We suppose that $\forall u \in [0,1]$, $x(1)$ and $y(u)$ are linkable and $x(u)$ and $y(0)$ are linkable. We can decompose $\eta^i$ as the following
\begin{equation}
\eta^i(y,x) = \eta^i_{s\mu}(y,x) dx^\mu + \eta^i_{t\mu}(y,x) dy^\mu
\end{equation}
where $\eta^i_s \in \underline{\Omega^1U^i_{(x)}}_{U^i_{(y)}}$ and $\eta^i_t \in \underline{\Omega^1U^i_{(y)}}_{U^i_{(x)}}$. By considering the three parts of $\mathcal C_{\gamma}^2$ we have
\begin{equation}
\oint_{\mathcal C_\gamma^2} \eta^i(y,x) = \int_{(y(0),x(0))}^{(y(1),x(1))} \eta^i(y,x) + \int_{y(1)}^{y(0)} \eta^i_t(y,x(1)) + \int_{x(1)}^{x(0)} \eta^i_s(y(0),x)
\end{equation}
By the Stokes theorem we have $\oint_{\mathcal C_\gamma^2} \eta^i = \iint_{\mathcal S_\gamma^2} \dd\eta^i = \iint_{\mathcal S_\gamma^2} B^i_{ns}$, and then
\begin{equation}
\int_{(y(0),x(0))}^{(y(1),x(1))} \eta^i(y,x) = \iint_{\mathcal S_{\gamma}^2} B^i_{ns}(y,x) + \int_{y(0)}^{y(1)} \eta^i_t(y,x(1)) + \int_{x(0)}^{x(1)} \eta^i_s(y(0),x)
\end{equation}
Finally the horizontal lift of $\gamma$ is associated with equation (\ref{hl2}).\\
We suppose now that $\gamma$ is impervious. By the property of $\eta^i$ with $t^{Lie}$ we have $t^{Lie}(\eta^i_s(y,x)) = -A^i(x)$ and $t^{Lie}(\eta^i_t(y,x)) = A^i(y)$, and then
\begin{equation}
t\left(e^{-\int_{s(\gamma)} \eta^i_s(y(0),x) - \int_{t(\gamma)}\eta^i_t(y,x(1))} \right) = \PP_{\mathcal C^1_\gamma}e^{-\oint_{\mathcal C^1_\gamma} A^i}
\end{equation}
By invoking a non-abelian Stokes theorem \cite{karp} we can write $\PP_{\mathcal C^1_\gamma}e^{-\oint_{\mathcal C^1_\gamma} A^i}$ as an ordering exponential along $\mathcal S^1_\gamma$ (a surface in $U^i$ having $\mathcal C^1_\gamma$ as boundary) of $F^i+t^{Lie}(B^i_{sph})$. But by construction $\PP_{\mathcal C^1_\gamma}e^{-\oint_{\mathcal C^1_\gamma} A^i} \in t(H)$ and then $\wp\left(\PP_{\mathcal C^1_\gamma}e^{-\oint_{\mathcal C^1_\gamma} A^i}\right) = e_{G/H}$. We can then choose $F^i = 0$ on $\mathcal S^1_\gamma$. We have then
\begin{eqnarray}
t\left(e^{-\int_{s(\gamma)} \eta^i_s(y(0),x) - \int_{t(\gamma)}\eta^i_t(y,x(1))} \right) & = & \PP_{\mathcal C^1_\gamma}e^{-\oint_{\mathcal C^1_\gamma} A^i}\\
& = & t\left(e^{- \iint_{\mathcal S^1_\gamma} B^i_{sph}} \right)
\end{eqnarray} 
It follows equation (\ref{hl3}).
\end{proof}

We have drastic reduction because we have supposed that $\forall u \in [0,1]$, $y(u)$ and $x(1)=y(1)$ are linkable and $x(u)$ and $x(0)=y(0)$ are linkable. If this is not the case, $\PP_{\mathcal C^1_\gamma}e^{-\oint_{\mathcal C^1_\gamma} A^i} \not\in t(H)$ ($\eta^i(y,x(1))$ and $\eta^i(y(0),x)$ are not defined).\\

For the general case (where $H$ is not the center of $G$ and is not necessary abelian) the situation is more complicated. 
\begin{property}
\label{hl0}
Let $\mathcal G$ be any Lie crossed module (not necessary a central extension of group). Let $\gamma$ be an elementary pseudosurface completely included in $\mathcal U^i$. The group element associated with the horizontal lift of $\gamma$ is
\begin{equation}
\PP_\gamma e^{-\int_{(y(0),x(0))}^{(y(u),x(u))} \underline \eta^i(y,x)} = \left( \PP_{\gamma}e^{-\int_{(y(0),x(0))}^{(y(u),x(u))} \eta^i(y,x)+\alpha^{Lie}_{A^i(x)}}, \PP_{s(\gamma)}e^{- \int_{x(0)}^{y(0)} A^i(x)} \right) \in H \rtimes G
\end{equation}
where $h_{\eta^i,A^i}(u) = \PP_{\gamma}e^{-\int_{(y(0),x(0))}^{(y(u),x(u))} \eta^i(y,x)+\alpha^{Lie}_{A^i(x)}} \in H$ is defined as the solution of the equation
\begin{equation}
\frac{dh_{\eta^i,A^i}}{du} = - \eta^i(y(u),x(u)) h_{\eta^i,A^i} - \alpha^{Lie}_{A^i(u)}\left(h_{\eta^i,A^i}\right)
\end{equation}
\end{property}

\begin{proof}
Let $U_{\eta+A} = \PP_\gamma e^{-\int_{(y(0),x(0))}^{(y(u),x(u))} \underline \eta^i(y,x)}$ and $U_A = \PP_{s(\gamma)}e^{- \int_{x(0)}^{y(0)} A^i(x)}$. Let $h_{\eta,A} = U_{\eta+A} U_A^{-1}$.
\begin{eqnarray}
\frac{dh_{\eta,A}}{du} & = & - (\eta+A) U_{\eta+A} U_A^{-1} + U_{\eta+A} U_A^{-1} A \\
& = & - \eta h_{\eta,A} - [A,h_{\eta,A}]
\end{eqnarray}
By the definition of the Lie braket $[A,h_{\eta,A}] = \alpha^{Lie}_A(h_{\eta,A})$. Because $\eta \in \Omega^1(U^i,\mathfrak h)$ and $\alpha^{Lie}_A \in \Omega^1(U^i,\mathrm{Der}(\mathfrak h))$, we have $h_{\eta,A} \in H$.
\end{proof}
\\

To make appear surface integrations for the element of $H$, we need use very complicated expressions issuing from the non-abelian Stokes theorem \cite{karp}. To avoid this difficulty we will consider an infinitesimal elementary pseudosurface. To this, we need some results of simplicial geometry \cite{albeverio1,albeverio2,beauce}. Let $\{K_n\}_{n \in \mathbb N}$ be a familly of smooth triangulations of $U^i \times U^i_{/ \mathcal R}$ (a triangulation is a triangular network covering $U^i \times U^i_{/ \mathcal R}$, a triangular cell of this network is called a simplex), such that $\bigcup_{n \in \mathbb N} K_n = U^i \times U^i_{/ \mathcal R}$. Let $(C^*(K_n,\mathfrak h),+,\cup,\delta)$ be the \v Cech differential algebra defined by:
\begin{itemize}
\item $C^p(K_n,\mathfrak h)$ is the algebra of antisymetric maps form $(K_n)^p$ to $\mathfrak h$ called the $p$-cochains.
\item $\delta: C^p(K_n,\mathfrak h) \to C^{p+1}(K_n,\mathfrak h)$ the cobord operator is $\forall \omega \in C^p(K_n,\mathfrak h)$
\begin{equation}
(\delta \omega)_{u_0...u_{p+1}} = \sum_{j=0}^{p+1} (-1)^j \omega_{u_0...\hat u_j...u_{p+1}}
\end{equation}
$u_i \in K_n$ and $\hat u_j$ signifies ``deprive of $u_j$''.
\item $\cup : C^p(K_n,\mathfrak h) \times C^q(K_n,\mathfrak h) \to C^{p+q}(K_n,\mathfrak h)$ the cup-product is $\forall \omega \in C^p(K_n,\mathfrak h)$, $\forall \eta \in C^q(K_n,\mathfrak h)$
\begin{equation}
(\omega \cup \eta)_{u_0...u_{p+q}} = \frac{1}{(p+1)!(q+1)!} \sum_{\sigma \in S_{p+q+1}} (-1)^\sigma [\omega_{u_{\sigma(0)}...u_{\sigma(p)}}, \eta_{u_{\sigma(p+1)}...u_{\sigma(p+q)}}]
\end{equation}
$S_{p+q+1}$ being the group of permutations and $(-1)^{\sigma}$ being the signature of the permutation $\sigma$.
\end{itemize}
At the inductive limit of the refinement $n \to +\infty$, the \v Cech differential algebra is isomorphic to the de Rham differential algebra $(\Omega^*(U^i \times U^i_{/ \mathcal R},\mathfrak h),+,\wedge,\dd)$. The isomorphism is induced by the de Rham map $R_n : \Omega^*(U^i \times U^i_{/\mathcal R},\mathfrak h) \to C^*(K_n,\mathfrak h)$, $\forall \omega \in \Omega^p(U^i \times U^i_{/ \mathcal R},\mathfrak h)$
\begin{equation}
R_n(\omega)_{u_0...u_p} = \int_{\langle u_0...u_p \rangle} \omega
\end{equation}
where $\langle u_0...u_p \rangle$ is a $p$ dimensional submanifold of $U^i \times U^i_{/ \mathcal R}$ forming a simplex with $u_0,...,u_p$ as vertices ($\langle u_0u_1 \rangle$ is an edge, $\langle u_0u_1u_2 \rangle$ is a triangular cell, $\langle u_0u_1u_2u_3 \rangle$ is a tetrahedron, etc). The reciprocal map is the Whitney map $W_n : C^*(K_n,\mathfrak h) \to \Omega^*(U^i \times U^i_{/\mathcal R},\mathfrak h)$ (see \cite{albeverio1,albeverio2,beauce}). It is interesting to note that $\forall \omega \in \Omega^p(U^i \times U^i_{/ \mathcal R},\mathfrak h)$
\begin{equation}
(\delta R_n(\omega))_{u_0...u_{p+1}} = \int_{\langle u_0...u_{p+1} \rangle} \dd \omega \iff \delta R_n(\omega) = R_n(\dd \omega)
\end{equation}
and $\forall \omega \in \Omega^p(U^i \times U^i_{/ \mathcal R})$, $\forall \eta \in \Omega^q(U^i \times U^i_{/ \mathcal R})$
\begin{equation}
\lim_{n \to +\infty} W_n(R_n(\omega) \cup R_n(\eta)) = \omega \wedge \eta
\end{equation}
(the limit being defined with the topology of a $L^2$-norm see \cite{albeverio1,albeverio2,beauce}). Let $\epsilon_n$ be ``the edge length'' of $K_n$ (i.e. $\forall \eta \in \Omega^1(U^i \times U^i_{/\mathcal R})$, $R_n(\eta)_{u_0u_1} = \mathcal O(\epsilon_n)$ with $\lim_{n \to + \infty} \epsilon_n = 0$). It is interesting to note that the Cartan structure equation $\beta = d\alpha + \alpha \wedge \alpha$ (with $\alpha \in \Omega^1(U^i \times U^i_{\mathcal R},\mathfrak h)$) takes the form
\begin{equation}
e^{R_n(\alpha)_{u_1u_2}}e^{-R_n(\alpha)_{u_0u_2}}e^{R_n(\alpha)_{u_0u_1}} = e^{R_n(\beta)_{u_0u_1u_2} + \mathcal O(\epsilon_n^3)}
\end{equation}
by using the Baker-Campbell-Hausdorff formula (\cite{BCH}) at the second order $e^a e^b = e^{a+b + \frac{1}{2}[a,b] + \mathcal O(\epsilon_n^3)}$ for $a,b \in \mathfrak h$ and $a,b=\mathcal O(\epsilon_n)$.\\
Let $\gamma : [0,1] \to \Morph(\mathcal M)$ be an elementary pseudosurface such that $(y(0),x(0))$, $(y(1),x(1))$, $(y(0),x(1)) \in K_n$ with $n$ large and such that $\forall u \in [0,1]$, $x(1)$ and $y(u)$ are linkable and $x(u)$ and $y(0)$ are linkable. Let $\mathcal C^2_\gamma$ be the closed path in $U^i \times U^i$ defined by  $[0,1] \ni u \mapsto (y(u),x(u))$ for its first part, $[0,1] \ni u \mapsto (y(1-u),x(1))$ for its second part, and $[0,1] \ni u \mapsto (y(0),x(1-u))$ for its last part. For the sake of simplicity we denote $u_0 = (y(0),x(0))$, $u_1=(y(1),x(1))$ and $u_2=(y(0),x(1))$. We can assimilate $\mathcal S^2_\gamma$ to the simplex (the triangular cell) $\langle u_0u_1u_2 \rangle$ of $K_n$. By using the Baker-Campbell-Hausdorff formula at the second order we have then
\begin{eqnarray}
& & e^{R_n(\eta^i)_{u_1u_2}} e^{-R_n(\eta^i)_{u_0u_2}} e^{-R_n(A^i)_{u_0u_2}} e^{R_n(\underline \eta^i)_{u_0 u_1}} \nonumber \\
& & = e^{\delta R_n(\eta^i)_{u_0u_1u_2} + R_n(\eta^i) \cup R_n(\eta^i)_{u_0u_1u_2} + R_n(\eta^i) \cup R_n(A^i)_{u_0u_1u_2} + R_n(A^i)\cup R_n(\eta^i)_{u_0u_1u_2} + \mathcal O(\epsilon_n^3)} \nonumber\\
& & = e^{R_n(B^i_{ns})_{u_0u_1u_2} + \mathcal O(\epsilon_n^3)}
\end{eqnarray}
Finally by writting that $\PP_\gamma e^{- \int_{(y(0),x(0))}^{(y(1),x(1))} \underline \eta^i(y,x)} \simeq e^{R_n(\underline \eta^i)_{u_0u_1}}$, the horizontal lift for an infinitesimal elementary pseudosurface $\gamma$ is
\begin{eqnarray}
& & \PP_\gamma e^{-\int_{(y(0),x(0))}^{(y(1),x(1))} \underline \eta^i(y,x)} \nonumber \\
& & \simeq \left(e^{-\iint_{\mathcal S_\gamma^2} B^i_{ns}(y,x)} e^{-\int_{s(\gamma)} \eta^i_s(y(0),x)} e^{- \int_{t(\gamma)} \eta^i_t(y,x(1))}, e^{-\int_{s(\gamma)} A^i(x)} \right)
\end{eqnarray}
Moreover, if $\gamma$ is impervious we have
\begin{eqnarray}
& & \PP_\gamma e^{-\int_{(y(0),x(0))}^{(y(1),x(1))} \underline \eta^i(y,x)} \nonumber \\
& & \simeq \left(e^{-\iint_{\mathcal S_\gamma^2} B^i_{ns}(y,x)} e^{- \iint_{\mathcal S_\gamma^1} B^i_{sph}(x)}, e^{-\int_{s(\gamma)} A^i(x)} \right)
\end{eqnarray}

\begin{definition}[Horizontal lift functor]
Let $\mathcal{H}\ell^i : \Morph(\mathcal U^i)^{[0,1]} \to H \rtimes G$ be the map which associates to an elementary pseudosurface the group element associated with its horizontal lift $\mathcal{H}\ell^i(\gamma) = \PP_\gamma e^{-\int_{(y(0),x(0))}^{(y(1),x(1))} \underline \eta^i(y,x)}$. We extend $\mathcal{H}\ell^i$ as a functor from $\mathcal{PS}(\mathcal U^i)$ to $\mathcal G$ transforming the horizontal compositions to horizontal compositions and the vertical compositions to vertical compositions:
\begin{eqnarray}
\mathcal H\ell^i (\gamma_1 * \gamma_2) & = & \mathcal H\ell^i(\gamma_1) \cdot \mathcal H\ell^i(\gamma_2) \\
\mathcal H\ell^i (\gamma_1 \circ \gamma_2) & = & \mathcal H\ell^i(\gamma_1) \circ \mathcal H\ell^i(\gamma_2)
\end{eqnarray}
``$\cdot$'' denotes the group law of $H \rtimes G$. 
\end{definition}
This functor permits to define the horizontal lift of any pseudosurface $\gamma$ of $\mathcal U^i$. Let a decomposition
\begin{equation}
\gamma = (\gamma_{11} \circ ... \circ \gamma_{1n}) * ... * (\gamma_{p1} \circ ... \circ \gamma_{pn})
\end{equation}
where each $\gamma_{ij}$ is an elementary pseudosurface. The horizontal lift of $\gamma$ is then
\begin{equation}
\mathcal H\ell^i(\gamma) = (\mathcal H\ell^i(\gamma_{11}) \circ ... \circ \mathcal H\ell^i(\gamma_{1n})) \cdot ... \cdot (\mathcal H\ell^i(\gamma_{p1}) \circ ... \circ \mathcal H\ell^i(\gamma_{pn}))
\end{equation}
This decomposition is well defined because of the exchange laws: $\forall \gamma_{11},\gamma_{12},\gamma_{21},\gamma_{22}\in \mathcal {PS}(\mathcal U^i)$ with $\Skel_{\gamma_{11}}(0) =\Skel_{\gamma_{21}}(1)$, $\Skel_{\gamma_{12}}(0)=\Skel_{\gamma_{22}}(1)$, $s(\gamma_{11}(u))=t(\gamma_{12}(u))$ and $s(\gamma_{21}(u))=t(\gamma_{22}(u))$
\begin{equation}
(\gamma_{11} \circ \gamma_{12}) * (\gamma_{21} \circ \gamma_{22}) = (\gamma_{11} * \gamma_{21}) \circ (\gamma_{12} * \gamma_{22})
\end{equation}
and $\forall h_{11},h_{12},h_{22},h_{21} \in H$, $\forall g_{12},g_{22} \in G$
\begin{eqnarray}
& & \left((h_{11},t(h_{12})g_{12}) \circ (h_{12},g_{12}) \right) \cdot \left((h_{21},t(h_{22})g_{22}) \circ (h_{22},g_{22}) \right) \nonumber \\
& & =  \left((h_{11},t(h_{12})g_{12})\cdot(h_{21},t(h_{22})g_{22})\right) \circ \left((h_{12},g_{12}) \cdot (h_{22},g_{22}) \right) 
\end{eqnarray}

\subsection{Horizontal lifts of pseudosurfaces crossing several 2-charts}
Now, we need to define the horizontal lifts for pseudosurfaces extending on several charts. Well defined horizontal lifts of paths and surfaces crossing several charts have been studied by Alvarez in \cite{alvarez}. Unfortunately, these results cannot be used directly in the present context.
\begin{proposition}
\label{compH}
Let $\gamma^i$ be an elementary pseudosurface of $\mathcal U^i$ and $\gamma^j$ be an elementary pseudosurface of $\mathcal U^j$ such that $\gamma^i(1) = \gamma^j(0)$. Let $x_\star = s(\gamma^i(1))=s(\gamma^j(0)) \in U^i \cap U^j$ and $y_\star=t(\gamma^i(1))=t(\gamma^j(0)) \in U^i \cap U^j$. Let $\gamma^{i\prime}$ and $\gamma^{j\prime}$ be two other elementary pseudosurfaces of $\mathcal U^i$ and $\mathcal U^j$ such that $\gamma^{j \prime} * \gamma^{i \prime} = \gamma^j * \gamma^i$ but with $(y_\star',x_\star') \not= (y_\star,x_\star)$. An horizontal lift of $\gamma^j * \gamma^i$ satisfying the condition
\begin{eqnarray}
\wp(s(\mathcal H\ell(\gamma^j*\gamma^i))) & = & \wp(s(\mathcal H\ell(\gamma^{j\prime}*\gamma^{i\prime}))) \\
\wp(t(\mathcal H\ell(\gamma^j*\gamma^i))) & = & \wp(t(\mathcal H\ell(\gamma^{j\prime}*\gamma^{i\prime})))
\end{eqnarray}
is defined by
\begin{equation}
\label{hlstar}
\mathcal H \ell(\gamma^j * \gamma^i) = \mathcal H\ell^j(\gamma^j) \cdot \left(h^{ij}(y_\star,x_\star),g^{ij}(x_\star)\right)^{-1} \cdot \mathcal H\ell^i(\gamma^i)
\end{equation}
\end{proposition}

\begin{proof}
We suppose that $\mathcal H \ell(\gamma^j * \gamma^i) = \mathcal H\ell^j(\gamma^j) \cdot q_\star \cdot \mathcal H\ell^i(\gamma^i)$ where $q_\star \in H \rtimes G$ is a transition element at $(y_\star,x_\star)$ between $\mathcal U^i$ and $\mathcal U^j$. We must have
\begin{equation}
\wp \left( s\left(\PP_{\gamma}e^{-\int_{u_\star}^{u_1} \underline \eta^j} \cdot q_\star \cdot \PP_{\gamma}e^{-\int_{u_0}^{u_\star} \underline \eta^i} \right) \right) = \wp \left( s\left(\PP_{\gamma}e^{-\int_{u_\star'}^{u_1} \underline \eta^j} \cdot q_\star' \cdot \PP_{\gamma}e^{-\int_{u_0}^{u_\star'} \underline \eta^i} \right) \right)
\end{equation}
where $u_\star = (y_\star,x_\star)$, $u_1 = (t(\gamma^j(1)),s(\gamma^j(1))) = (t(\gamma^{j\prime}(1)),s(\gamma^{j\prime}(1)))$ and $u_0 = (t(\gamma^i(0)),s(\gamma^i(0))) = (t(\gamma^{i\prime}(0)),s(\gamma^{i\prime}(0)))$ (and $\gamma = \gamma^{j \prime} * \gamma^{i \prime} = \gamma^j * \gamma^i$). Since $\left(\PP_{\gamma}e^{-\int_{u_\star'}^{u_1} \underline \eta^j}\right)^{-1} \PP_{\gamma}e^{-\int_{u_\star}^{u_1} \underline \eta^j} = \PP_{\gamma}e^{-\int_{u_\star}^{u_\star'} \underline \eta^j}$, we have
\begin{equation}
\wp \left( s\left( \PP_\gamma e^{-\int_{u_\star}^{u_\star'} \underline \eta^j} \cdot q_\star \right) \right) = \wp\left(s\left(q_\star' \cdot \PP_\gamma e^{-\int_{u_\star}^{u_\star'} \underline \eta^i} \right) \right)
\end{equation}
Because of the gluing relation of $\underline \eta$, and by an argument similar to the property \ref{hl0} we have
\begin{eqnarray}
s\left(\PP_\gamma e^{-\int_{u_\star}^{u_\star'} \underline \eta^j}\right) & = & h_\star s\left(\PP_\gamma e^{\int_{u_\star}^{u_\star'} q^{ij-1} \underline \eta^i q^{ij} + q^{ij-1} d_{(2)} q^{ij}} \right) \\
& = & h_\star s\left( q^{ij}(u_\star')^{-1} \PP_\gamma e^{\int_{u_\star}^{u_\star'} \underline \eta^i} q^{ij}(u_\star) \right)
\end{eqnarray}
where $h_\star= \PP_\gamma e^{-\int_{u_\star}^{u_\star'} t^{Lie}(\eta^{ij}(x)) + \alpha^{Lie}_{t^{Lie}_\bullet(q^{ij-1} \underline \eta^i q^{ij} + q^{ij-1} d_{(2)} q^{ij})}} \in H$. We have then
\begin{equation}
\wp\left(s\left(q^{ij}(u_\star')^{-1} \PP_\gamma e^{\int_{u_\star}^{u_\star'} \underline \eta^i} q^{ij}(u_\star) \cdot q_\star \right) \right) = \wp\left(s\left(q_\star' \cdot \PP_\gamma e^{-\int_{u_\star}^{u_\star'} \underline \eta^i} \right) \right)
\end{equation}
and then
\begin{equation}
\wp\left(s\left(\PP_\gamma e^{\int_{u_\star}^{u_\star'} \underline \eta^i} \cdot q^{ij}(u_\star) q_\star \right) \right) = \wp\left(s\left( q^{ij}(u_\star')^{-1} q_\star' \cdot \PP_\gamma e^{-\int_{u_\star}^{u_\star'} \underline \eta^i} \right) \right)
\end{equation}
It follows that $q_\star = q^{ij}(u_\star)^{-1}$ (modulo an $H$ element without significance). Since the calculus is the same for the target condition ($h_\star= \PP_\gamma e^{-\int_{u_\star}^{u_\star'} t^{Lie}(\eta^{ij}(y)) + \alpha^{Lie}_{t^{Lie}_\bullet(q^{ij-1} \underline \eta^i q^{ij} + q^{ij-1} d_{(2)} q^{ij})}}$), we dot not have another result.
\end{proof}

It is important to note that except for trivial 2-bundles (where $\eta^{ij} = 0$), $\mathcal H\ell(\gamma^j * \gamma^i) \not= \mathcal H \ell(\gamma^{j\prime} * \gamma^{i\prime})$, and the horizontal lift of a pseudosurface extending on two charts depends on an arbitrary point $(y_\star,x_\star)$ chosen for the transition. This is a consequence of the impossibility to lift $\{P^i\}_i$ and $\{Q^i\}_i$ to usual bundles. The consistency of the connective structure being defined by the global $G/t(H)$-bundle $R$, it is natural that the consistency of $\mathcal H \ell (\gamma^j * \gamma^i)$ is ensured only for its projection by $\wp : G \to G/t(H)$.\\

\begin{proposition}
\label{compV}
Let $\gamma^i$ be an elementary pseudosurface of $\mathcal U^i$ and $\gamma^j$ be an elementary pseudosurface of $\mathcal U^j$ such that $s(\gamma^j) = t(\gamma^i) = \mathcal C_\star \subset U^i \cap U^j$. An horizontal lift of $\gamma^j \circ \gamma^i$ permitting the composition of $\mathcal H\ell^j(\gamma^j)$ and $\mathcal H\ell^i(\gamma^i)$ is defined by
\begin{eqnarray}
\mathcal H\ell(\gamma^j \circ \gamma^i) & = &\mathcal H\ell^j(\gamma^j) \circ \nonumber \\
& & \quad \left( (e_H,g^{ij}(x_\star(1))^{-1})\cdot \right. \nonumber \\
& & \qquad \left(\left(\PP_{\mathcal C_\star}e^{-\int_{\mathcal C_\star} \alpha_{g^{ij}}(\eta^{ij}) + \alpha^{Lie}_{A^i}},\PP_{\mathcal C_\star} e^{- \int_{\mathcal C_\star} A^i}\right) \circ \mathcal H\ell^i(\gamma^i)\right) \nonumber \\
& & \quad \left. \cdot(e_H,g^{ij}(x_\star(0))) \right)
\end{eqnarray}
with $s(\gamma^j(u))=t(\gamma^i(u))=x_\star(u)$ ($\forall u \in [0,1]$).
\end{proposition}

\begin{proof}
The vertical composition of $\mathcal H \ell^i(\gamma^i)$ with $\mathcal H \ell^j(\gamma^j)$ cannot directly be performed since
\begin{equation}
t(\mathcal H\ell^i(\gamma^i)) = \PP_{\mathcal C_\star} e^{- \int_{\mathcal C_\star} A^i} \not= \PP_{\mathcal C_\star} e^{- \int_{\mathcal C_\star} A^j} = s(\mathcal H\ell^j(\gamma^j))
\end{equation}
Since $A^j = g^{ij-1} A^i g^{ij} + g^{ij-1}dg^{ij} + t^{Lie}(\eta^{ij})$ we have
\begin{equation}
\PP_{\mathcal C_\star} e^{- \int_{\mathcal C_\star} A^j} = g^{ij}(x_\star(1))^{-1} \PP_{\mathcal C_\star} e^{- \int_{\mathcal C_\star} A^i + t^{Lie}(\alpha_{g^{ij}}(\eta^{ij}))} g^{ij}(x_\star(0)) 
\end{equation}
Since we have
\begin{equation}
s\left(\PP_{\mathcal C_\star}e^{-\int_{\mathcal C_\star} \alpha_{g^{ij}}(\eta^{ij}) + \alpha^{Lie}_{A^i}},\PP_{\mathcal C_\star} e^{- \int_{\mathcal C_\star} A^i} \right) = \PP_{\mathcal C_\star} e^{- \int_{\mathcal C_\star} A^i}
\end{equation}
and since by using an argument similar to the property \ref{hl0} we have
\begin{equation}
t\left(\PP_{\mathcal C_\star}e^{-\int_{\mathcal C_\star} \alpha_{g^{ij}}(\eta^{ij}) + \alpha^{Lie}_{A^i}},\PP_{\mathcal C_\star} e^{- \int_{\mathcal C_\star} A^i} \right)  = \PP_{\mathcal C_\star} e^{- \int_{\mathcal C_\star} A^i + t^{Lie}(\alpha_{g^{ij}}(\eta^{ij}))} 
\end{equation}
we conclude that the different arrows can be composed.
\end{proof}

In contrast to the horizontal composition, $\mathcal H\ell(\gamma^j \circ \gamma^i)$ as defined by this proposition does not depend on an arbitray choice. $\mathcal C_\star$ and its end points are fixed by the source and target maps.

\begin{proposition}
Let $\gamma^i$, $\gamma^j$, and $\gamma^k$ be elementary pseudosurfaces of $\mathcal U^i$, $\mathcal U^j$ and $\mathcal U^k$ such that $s(\gamma^k)=t(\gamma^j)=\mathcal C_{\star1} \in U^k \cap U^j$, $s(\gamma^i(1))=s(\gamma^j(0))=t(\gamma^j(0))=s(\gamma^k(0))=x_\star \in U^i \cap U^j \cap U^k$ and $t(\gamma^i(1))=t(\gamma^k(0)) = y_\star \in U^i \cap U^k$. An horizontal lift of $(\gamma^k \circ \gamma^j) * \gamma^i$ compatible with the previous definitions and with the exchange law:
\begin{equation}
\mathcal H\ell\left((\gamma^k \circ \gamma^j\right) * \gamma^i) = \mathcal H\ell\left((\gamma^k * \gamma^i) \circ (\gamma^j * \id_{s(\gamma^i)}) \right)
\end{equation}
is defined by
\begin{eqnarray}
\mathcal H\ell\left((\gamma^k \circ \gamma^j\right) * \gamma^i) & = & \mathcal H\ell(\gamma^k \circ \gamma^j) \cdot \nonumber \\
& & \left\{(h^{ik}(y_\star,x_\star),g^{ik}(x_\star))^{-1}\mathcal H\ell^i(\gamma^i) \circ \right. \nonumber \\
& &  \left. (h^{ijk}(x_\star),g^{ij}(x_\star)g^{jk}(x_\star))^{-1} \PP_{s(\gamma^i)}e^{-\int_{s(\gamma^i)}A^i} t(k^i(x_0)) \right\}
\end{eqnarray}
with $x_0 = s(\gamma^i(0))$. \textit{To simplify the notations we have denoted $(e_H,g)$ by $g$ and we have supposed that the horizontal composition ``$\cdot$'' precedes the vertical composition ``$\circ$''.} 
\end{proposition}

\begin{proof}
By applying the exchange law we have $(\gamma^k \circ \gamma^j) * \gamma^i = (\gamma^k \circ \gamma^j) * (\gamma^i \circ \id_{s(\gamma^i)}) = (\gamma^k * \gamma^i) \circ (\gamma^j * \id_{s(\gamma^i)})$. By applying the proposition \ref{compV} we have
\begin{equation}
\label{dblcompo}
\mathcal H\ell((\gamma^k*\gamma^i)\circ(\gamma^j*\id_{s(\gamma^i)}))=\mathcal H\ell(\gamma^k*\gamma^i)\circ\left\{g^{jk}(x_1)^{-1}(q(\mathcal C_\star)\circ \mathcal H\ell(\gamma^j*\id_{s(\gamma^i)}))t(k^i(x_0)) \right\}
\end{equation}
with $x_1 = s(\gamma^k(1))=t(\gamma^j(1))$, $\mathcal C_\star = \mathcal C_{\star1} \cup s(\gamma^i)$ and $q(\mathcal C_\star)$ is defined as in the proposition \ref{compV}. More precisely, by using the gluing relations for $A^j$ and for $\eta^{jk}$ we find that
\begin{eqnarray}
A^j+t^{Lie}(\alpha_{g^{jk}}(\eta^{jk})) & = & g^{ij-1}A^ig^{ij}+g^{ij-1}dg^{ij} \nonumber \\
& & + g^{ij-1} t^{Lie}\left(\alpha_{g^{ij}}(\eta^{ij})+\alpha_{g^{ij}g^{jk}}(\eta^{jk})\right) g^{ij} \\
& = & g^{ij-1}A^ig^{ij}+g^{ij-1}dg^{ij} \nonumber \\
& &  + g^{ij-1} t^{Lie}\left(h^{ijk-1}dh^{ijk}+h^{ijk-1}\alpha^{Lie}_{A^i}(h^{ijk}) \right. \nonumber \\
& & \qquad \left. +h^{ijk-1}\alpha_{g^{ik}}(\eta^{ik})h^{ijk}\right) g^{ij} \\
& = & g^{ij-1}t(h^{ijk})^{-1} \left(A^i+t^{Lie}(\alpha_{g^{ik}}(\eta^{ik})) \right)t(h^{ijk})g^{ij} \nonumber \\
& & \quad +g^{ij-1}t(h^{ijk})^{-1}d(h^{ijk}g^{ij})
\end{eqnarray}
It follows that
\begin{equation}
t(q(\mathcal C_\star)) = \PP_{\mathcal C_{\star1}}e^{-\int_{\mathcal C_{\star1}}A^j+t^{Lie}(\alpha_{g^{jk}}(\eta^{jk}))}g^{ij}(x_\star)^{-1}t(h^{ijk}(x_\star))^{-1} \PP_{s(\gamma^i)}e^{-\int_{s(\gamma^i)} A^i}
\end{equation}
and then
\begin{equation}
q(\mathcal C_\star) = q^{jk}(\mathcal C_{\star1}) (h^{ijk}(x_\star),g^{ij}(x_\star))^{-1} \PP_{s(\gamma^i)}e^{-\int_{s(\gamma^i)} A^i}
\end{equation}
We have then
\begin{eqnarray}
& & g^{jk}(x_1)^{-1}(q(\mathcal C_\star)\circ \mathcal H\ell(\gamma^j*\id_{s(\gamma^i)}))t(k^i(x_0))  \nonumber \\
& & = \left\{g^{jk}(x_1)^{-1}q^{jk}(\mathcal C_{\star1}) g^{jk}(x_\star)\right\} \cdot \left\{(h^{ijk}(x_\star),g^{ij}(x_\star)g^{jk}(x_\star))^{-1}\PP_{s(\gamma^i)}e^{-\int_{s(\gamma^i)} A^i} t(k^i(x_0)) \right\} \circ \nonumber \\
& & \quad \left\{g^{jk}(x_1)^{-1}\mathcal H\ell^k(\gamma^k) g^{jk}(x_\star) \right\} \cdot \left\{g^{jk}(x_\star)^{-1}g^{ij}(x_\star)^{-1} \PP_{s(\gamma^i)}e^{-\int_{s(\gamma^i)} A^i} t(k^i(x_0)) \right\}
\end{eqnarray}
By using the exchange law for the elements delimited by $\{ \}$ in the previous expression, we find
\begin{eqnarray}
& & g^{jk}(x_1)^{-1}(q(\mathcal C_\star)\circ \mathcal H\ell(\gamma^j*\id_{s(\gamma^i)}))t(k^i(x_0))  \nonumber \\
& & = \left\{g^{jk}(x_1)^{-1}(q^{jk}(\mathcal C_{\star 1}) \circ \mathcal H\ell^j(\gamma^j))g^{jk}(x_\star) \right\} \cdot \nonumber \\
& & \quad \left\{(h^{ijk}(x_\star),g^{ij}(x_\star)g^{jk}(x_\star))^{-1} \PP_{s(\gamma^i)}e^{-\int_{s(\gamma^i)} A^i} t(k^i(x_0)) \right\}
\end{eqnarray}
Since $\mathcal H\ell(\gamma^k * \gamma^i) = \mathcal H\ell^k(\gamma^k)(h^{ik}(y_\star,x_\star),g^{ik}(x_\star))^{-1} \mathcal H\ell^i(\gamma^i)$, equation (\ref{dblcompo}) becomes
\begin{eqnarray}
& & \mathcal H\ell((\gamma^k*\gamma^i)\circ(\gamma^j*\id_{s(\gamma^i)})) \nonumber \\
& & = \left\{\mathcal H\ell^k(\gamma^k)\right\} \cdot \left\{(h^{ik}(y_\star,x_\star),g^{ik}(x_\star))^{-1} \mathcal H\ell^i(\gamma^i) \right\} \circ \nonumber \\
& & \quad \left\{g^{jk}(x_1)^{-1}(q^{jk}(\mathcal C_{\star 1}) \circ \mathcal H\ell^j(\gamma^j))g^{jk}(x_\star) \right\} \cdot \nonumber \\
& & \quad \left\{(h^{ijk}(x_\star),g^{ij}(x_\star)g^{jk}(x_\star))^{-1} \PP_{s(\gamma^i)}e^{-\int_{s(\gamma^i)} A^i} t(k^i(x_0)) \right\}
\end{eqnarray}
By using the exchange law for the elements delimited by $\{ \}$ we find the result of the proposition.
\end{proof}

Similar formulae for the other situations with a branching point on a triple overlap can be obtained by the same manner.\\

With all these definitions, $\mathcal H\ell$ can be extended as a functor of $\bigsqcup_i \mathcal{PS}(\mathcal U^i)$ to $\mathcal G$ and define any horizontal lift relative to a 2-cover $\{\mathcal U_i\}_i$. $\mathcal H\ell$ can be extended as a functor of $\mathcal{PS}(\mathcal M)$ to $\mathcal G$ only if the 2-bundle is trivial (independent from the choice of a 2-cover and then on the choices of transition points for pseudosurfaces extending on several 2-charts). 

\section{Example: The Bloch wave operators in quantum dynamics}
\subsection{The Bloch wave operators}
The studies of quantum dynamical systems with ``large Hilbert space'', i.e. quantum systems involving a large number of independent states in their dynamics, are generally difficult for the theoretical viewpoint as for the numerical viewpoint. Methods involving active spaces, effective hamiltonians and wave operators are good tools to solve this problem (see \cite{killingbeck,jolicard,durand}).\\
Let $G_m(\mathcal H) = \{P \in \mathcal B(\mathcal H), P^2=P, P^\dagger=P, \tr P = m \}$ be the space of rank $m$ orthogonal projectors of the separable Hilbert space $\mathcal H$ ($\mathcal B(\mathcal H)$ denotes the set of bounded operators of $\mathcal H$). If $\mathcal H$ is finite dimensional, i.e. $\mathcal H \simeq \mathbb C^n$, $G_m(\mathbb C^n)$ is a complex manifold called a complex grassmanian \cite{rohlin}. This manifold is endowed with a K\"ahlerian structure (see \cite{nakahara}), and particularly with a distance (called the Fubini-Study distance) defined by
\begin{equation}
\forall P_1,P_2 \in G_m(\mathbb C^n), \quad \dist(P_1,P_2) = \arccos|\det Z_1^\dagger Z_2|^2
\end{equation}
where $Z_1, Z_2 \in \mathfrak M_{n \times m}(\mathbb C)$ are the matrices of two arbitrary orthonormal basis of $\Ran P_1$ and $\Ran P_2$ expressed in an orthonormal basis of $\mathbb C^n$. We can note that $0 \leq \dist(P_1,P_2) \leq \frac{\pi}{2}$. The Fubini-Study distance measures the ``quantum compatibility'' between the two subspaces $\Ran P_1$ and $\Ran P_2$ in the sense that $\dist(P_1,P_2) = \frac{\pi}{2}$ if and only if $\Ran P_1^\bot \cap \Ran P_2 \not=\{0\}$ or $\Ran P_1 \cap \Ran P_2^\bot \not=\{0\}$, i.e. there exists a state of $\Ran P_1$ for which the probability of obtaining the same measures as that with a system in a state of $\Ran P_2$ is zero (see \cite{viennot6}). For infinite dimensional Hilbert space, it is possible to define a manifold $G_m(\mathcal H)$ endowed with a K\"ahlerian structure by using the inductive limit technique (see \cite{rohlin}).\\
Let $P_0,P \in G_m(\mathcal H)$ be such that $\dist(P_0,P) < \frac{\pi}{2}$. We call wave operator associated with $\Ran P_0$ and $\Ran P$ the operator $\Omega \in \mathcal B(\mathcal H)$ defined by
\begin{equation}
\Omega = P(P_0PP_0)^{-1}
\end{equation}
where $(P_0PP_0)^{-1} = P_0(P_0PP_0)^{-1} P_0$ is the inverse of $P$ within $\Ran P_0$ (it exists only if $P$ is not too far from $P_0$, i.e. $\dist(P,P_0) < \frac{\pi}{2}$). Usually the wave operators are used to solve eigenequation \cite{killingbeck}. In that case, we solve an effective eigenequation $H^{eff} \psi_0 = \lambda \psi_0$ where $H^{eff} = P_0 H \Omega \in \mathcal L(\Ran P_0)$ is the effective Hamiltonian within $\Ran P_0$ ($H \in \mathcal B(\mathcal H)$ is the true self-adjoint Hamiltonian), and we recover the true eigenvector associated with $\lambda$, $H\psi = \lambda \psi$, by $\psi = \Omega \psi_0 \in \Ran P$ ($\psi_0 = P_0 \psi$). $\Omega$ is called a Bloch wave operator and is obtained by solving the Bloch equation
\begin{equation}
[H,\Omega] \Omega = 0
\end{equation}
Since $\Omega^2 = \Omega$, a Bloch wave operator can be viewed as a non-linear generalization of an eigenprojector (an eigenprojector satisfying $[H,P] = 0$ with $P^2=P$). Physically, a Bloch wave operator compares the approximate eigenstates within $\Ran P_0$ (which is called the active subspace) with the associated true eigenstates.\\
We can define a weak left inverse of a wave operator: if $\Omega = P(P_0PP_0)^{-1}$ then $\Omega^{-1} = P_0P$ satisfies $\Omega^{-1} \Omega = P_0$.\\

In a same manner, in order to compare an approximate quantum dynamics within an active space $\Ran P_0$ with the true dynamics, we can introduce the time-dependent wave operator \cite{jolicard} :
\begin{equation}
\Omega(t) = P(t)(P_0 P(t) P_0)^{-1}
\end{equation}
where $(P_0 P(t) P_0)^{-1}$ is still the inverse within $\Ran P_0$, and where $t \mapsto P(t) \in G_m(\mathcal H)$ is the solution of the Schr\"odinger-von Neumann equation :
\begin{equation}
\ihbar \dot P(t) = [H(t),P(t)] \qquad P(0) = P_0
\end{equation}
$H(t) \in \mathcal B(\mathcal H)$ being the self-adjoint time-dependent Hamiltonian. We can then solve the effective Schr\"odinger equation within $\Ran P_0$, $\ihbar \partial_t \psi_0(t) = H^{eff}(t)\psi_0(t)$, where $H^{eff}(t) = P_0 H(t) \Omega(t) \in \mathcal L(\Ran P_0)$ is the effective Hamiltonian, and recover the true wave function, $\ihbar \partial_t \psi(t) = H(t) \psi(t)$, by $\psi(t) = \Omega(t)\psi_0(t)$ ($P_0 \psi(t) = \psi_0(t)$). The time-depend wave operator can be used only if the dynamics does not escape too far from the initial subspace, i.e. $\forall t$, $\dist(P(t),P_0) < \frac{\pi}{2}$. Since $P(t) = U(t,0) P_0 U(t,0)^\dagger$, where $U(t,0) \in \mathcal U(\mathcal H)$ is the evolution operator ($\ihbar \dot U(t,0) = H(t)U(t,0)$, $U(0,0)= 1$; $\mathcal U(\mathcal H)$ denotes the set of unitary operators of $\mathcal H$), we can also write
\begin{equation}
\Omega(t) = U(t,0)(P_0U(t,0)P_0)^{-1}
\end{equation}
By using this expression, it is not difficult to prove that the time-dependent wave operator satisfies
\begin{equation}
\ihbar \dot \Omega(t) = [H(t),\Omega(t)]\Omega(t) \qquad \Omega(0)=P_0
\end{equation}

We can also introduce the generalized time-dependent wave operator \cite{viennot7}:
\begin{equation}
\Omega(t) = P(t)(P_0(t)P(t)P_0(t))^{-1}
\end{equation}
where $t \mapsto P(t) \in G_m(\mathcal H)$ is the solution of the Schr\"odinger-von Neumann equation and where $t \mapsto P_0(t) \in G_m(\mathcal H)$ is a $\mathcal C^2$ instantaneous eigenprojector: $\forall t$, $[H(t),P_0(t)]=0$. This wave operator satisfies
\begin{equation}
\ihbar \dot \Omega(t) = [H(t),\Omega(t)] \Omega(t) + \ihbar \Omega(t)\dot \Omega(t) \qquad \Omega(0) = P_0(0)
\end{equation}
This wave operator can be used to treat an almost adiabatic dynamics where the dynamics does not escape too far from the instantaneous eigenspace, i.e. $\forall t$, $\dist(P(t),P_0(t)) < \frac{\pi}{2}$. Let $H^{eff}(t) = \Omega(t)^{-1} H(t) \Omega(t) \in \mathcal L(\Ran P_0(t))$ be an effective hamiltonian within $\Ran P_0(t)$. Let $\{\phi_{0a}(t) \in \Ran P_0(t)\}_{a=1,...,m}$ be a complete set of eigenvectors of $H^{eff}(t)$ (for the sake of simplicity, we consider here that $H^{eff}$ is diagonalizable), associated with the eigenvalues $\{\lambda_a^{eff}(t)\}_{a=1,...,m}$. Let $\psi(t)$ be the true wave function which is the solution of the Schr\"odinger equation $\ihbar \partial_t \psi(t) = H(t)\psi(t)$ with $\psi(0) = \phi_{0a}(0)$. Since $\dist(P(t),P_0(t)) < \frac{\pi}{2}$ we wan write that $\psi(t) = \sum_{b=1}^m c_b(t) \Omega(t)\phi_{0b}(t)$. By injected this expression in the Schr\"odinger equation, we find that
\begin{equation}
\label{wowf}
\psi(t) = \sum_{b=1}^m \left[ \Te^{-\ihbar^{-1} \int_0^t E^{eff}(t')dt'- \int_0^t A(t')dt' - \int_0^t \eta(t')dt'} \right]_{ba} \Omega(t) \phi_{0b}(t)
\end{equation}
where $\Te$ is the time ordering exponential (the Dyson series) and where the matrices $E^{eff}$, $A$, $\eta \in \mathfrak{M}_{m \times m}(\mathbb C)$ are defined by
\begin{eqnarray}
E^{eff}(t) & = & \mathrm{diag}(\lambda_1^{eff}(t),...,\lambda_m^{eff}(t)) \\
A(t) & = & (Z_0(t)^\dagger Z_0(t))^{-1} Z_0(t)^\dagger \partial_t Z_0(t) \\
\eta(t) & = & (Z_0(t)^\dagger Z_0(t))^{-1} Z_0(t)^\dagger \Omega(t)^{-1} \dot \Omega(t) Z_0(t)
\end{eqnarray}
where $Z_0(t) \in \mathfrak M_{n \times m}(\mathbb C)$ is the matrix representing $(\phi_{01}(t),...,\phi_{0m}(t))$ in a fixed orthonormal basis of $\mathcal H \simeq \mathbb C^n$ (if $\mathcal H$ is infinite dimensional, $Z_0(t) \in \left(\ell^2(\mathbb N)\right)^{\otimes m}$, $\ell^2(\mathbb N)$ denoting the square integrable sequences representing the coefficients of the decomposition of the states of $\mathcal H$ on a fixed orthonormal basis). $A$ and $\eta$ are the generators of two kinds of non-abelian geometric phases. The next section discusses the geometric structure in which they take place.\\
Remark: the geometric structure associated with usual time-dependent wave operators $P(t)(P_0P(t)P_0)^{-1}$ (with $\dot P_0 = 0$) has been studied in \cite{viennot6}. The present work focus on the generalized time-dependent wave operators ($\dot P_0 \not= 0$).

\subsection{The category of the $m$-dimensional subspaces}
Before introducing the affine 2-space of the wave operators, we need to introduce an intermediate category.\\
We denote by $\mathcal L_m^\infty(\mathcal H)$ the set of rank $m$ linear operators of $\mathcal H$. For an endomorphism $f \in \mathcal L_m^\infty(\mathcal H)$ we consider the decomposition $\ker f \overset{\bot}{\oplus} \ker f^\bot$ where $\dim \ker f^\bot = \dim \Ran f = m$. We introduce moreover the set 
$$ \mathcal L_m^1(\mathcal H) = \{ f \in \mathcal L_m^\infty(\mathcal H), \dist(\ker f^\bot,\Ran f) < \frac{\pi}{2} \}$$
and $\forall q \in \mathbb N^*$ we set
$$ \mathcal L_m^q(\mathcal H) = \{(f_q,...,f_1) \in (\mathcal L_m^1(\mathcal H))^q, \Ran f_i = \ker f_{i+1}^\bot\}/\chi $$
where the equivalence relation is defined by
$$ (f_q,...,f_1) \sim_\chi (f'_q,...,f'_1) \iff \begin{cases} \chi(f_q,...,f_1) = \chi(f'_q,...,f'_1) & \\ \Ran f_i = \Ran f_i', \forall i & \end{cases} $$
with $\chi(f_q,...,f_1) = f_q...f_1$ (the products being the operator composition).\\
 
Let $\mathcal E$ be the category defined by
\begin{itemize}
\item $\Obj(\mathcal E)$ are the $m$-dimensional vector subspaces of $\mathcal H$.
\item $\Morph(\mathcal E) = \bigsqcup_{q=1}^\infty \mathcal L_m^q(\mathcal H)$ (we note that $\chi(\Morph(\mathcal E)) = \mathcal L_m^\infty(\mathcal H)$).
\item $\forall E \in \Obj(\mathcal E)$, $\id_E = P_E$ (the orthogonal projection on $E$).
\item $\forall f \in \Morph(\mathcal E)$, $s(f) = \ker \chi(f)^\bot$ and $t(f) = \Ran \chi(f)$.
\item $\forall f,g \in \Morph(\mathcal E)$, $\Ran \chi(g) = \ker \chi(f)^\bot$; $g \circ f = [g_q,...,g_1,f_p,...,f_1]_{\chi}$ where $g=[g_q,...,g_1]_{\chi}$, $f=[f_p,...,f_1]_{\chi}$, $[.]_{\chi}$ denoting the equivalence class associated with $\sim_\chi$.
\end{itemize}

\subsection{The affine 2-space of the wave operators}
Let $(\mathcal M,\mathcal R,\Omega)$ be the hyperbolic affine 2-space defined by
\begin{itemize}
\item $\Obj(\mathcal M) = G_m(\mathcal H)$.
\item $\forall P,Q \in G_m(\mathcal H)$, $P \mathcal R Q \iff \dist(P,Q) < \frac{\pi}{2}$.
\item $\Morph(\mathcal M) = \{P_q(P_{q-1}P_qP_{q-1})^{-1}...(P_2P_1P_2)^{-1}, P_i \in G_m(\mathcal H), \dist(P_{i+1},P_i) < \frac{\pi}{2} \}$
\item $\Omega(P_q,...,P_1) = P_q(P_{q-1}P_qP_{q-1})^{-1}...(P_2P_1P_2)^{-1} = \overleftarrow{P_q...P_1}$.
\item $s(P_q(P_{q-1}P_qP_{q-1})^{-1}...(P_2P_1P_2)^{-1}) = \Ran P_1$, $t(P_q(P_{q-1}P_qP_{q-1})^{-1}...(P_2P_1P_2)^{-1}) = \Ran P_q$, $\id_{P} = P(PPP)^{-1} = P$.
\item The arrow composition is just the operator composition applied on the wave operators.
\end{itemize}
Let $\varpi \in \Funct(\mathcal E,\mathcal M)$ be the functor consisting to transform the vector spaces into their orthogonal projectors, and such that
\begin{equation}
\varpi([f_q,...,f_1]_{\chi}) = P_{\Ran f_q}(P_{\Ran f_{q-1}}P_{\Ran f_q} P_{\Ran f_{q-1}})^{-1}...(P_{\ker f_1^\bot} P_{\Ran f_1} P_{\ker f_1^\bot})^{-1}
\end{equation}
$P_{\Ran f_i}$ being the orthogonal projector on $\Ran f_i$.\\

Let $(e_a)_{a=1,...,n}$ be the chosen orthonormal basis of $\mathcal H$. Let $\{P^i\}_i$ be the set of orthogonal projectors on the spaces spaned by $m$ vectors of $(e_a)_{a=1,...,n}$. We denote by $I^i$ the set of indices of the $m$ vectors spaning $\Ran P^i$ ($\Ran P^i = \Span(e_a;a\in I^i)$). Let $U^i$ be the open chart of $G_m(\mathbb C^n)$ defined by
\begin{equation}
U^i = \{P \in G_m(\mathbb C^n) | \dist(P,P^i) < \frac{\pi}{2} \}
\end{equation}
$\{U^i\}_i$ constitues a good open cover of $G_m(\mathbb C^n)$ and then $\{\mathcal U^i \}_i$ generated by $\Omega$ constitutes a good open 2-cover of $\mathcal M$. $\forall P \in U^i$, there exists a basis $(u_a)_{a \in I^i}$ of $\Ran P$ such that (see \cite{rohlin}):
\begin{equation}
u_a = e_a + \sum_{b \not\in I^i} c_{ab} e_b \qquad c_{ab} \in \mathbb C
\end{equation}
The map $\xi^i : \begin{array}{rcl} U^i & \to & \mathbb C^{m(n-m)} \\ P & \mapsto & (c_{ab})_{a \in I^i, b \not\in I^i} \end{array}$ is the coordinates map of $U^i$. $\forall P \in U^i$ we denote by $Z^i_0 \in \mathfrak M_{n \times m}(\mathbb C)$ the matrix representing $(u_1,...,u_m)$ in the basis $(e_a)_{a=1,...,n}$ (we call it the coordinates matrix of $P$).

\subsection{The trivial 2-bundle associated with the wave operators}
Let $\mathcal P$ be the category defined by
\begin{itemize}
\item $\Obj(\mathcal P) = \{Z \in \mathfrak M_{n \times m}(\mathbb C), \det(Z^\dagger Z) \not= 0 \}$. $\Obj(\mathcal P)$ can be identified with the complex non-compact Stiefel manifold (see \cite{rohlin}).
\item $\Morph(\mathcal P) = \{(f,Z) \in \Morph(\mathcal E) \times \Obj(\mathcal P); s(f) = \Span(Z) \}$. ($\Span(Z)$ denotes the vector space spanned by the vectors represented by $Z$).
\item $s(f,Z) = Z$; $t(f,Z)=\chi(f)Z$ ($\chi(f)Z$ denotes the matrix in the fixed basis $(e_a)_{a=1,...,n}$ representing the action of $\chi(f)$ on the $m$ vectors represented by $Z$).
\item $\id_{Z} = (P,Z)$ where $P = Z(Z^\dagger Z)^{-1}Z^\dagger$ is the orthogonal projector on $\Span(Z)$.
\item $(f_2,W) \circ (f_1,Z) = (f_2 \circ f_1,Z)$ with $W = \chi(f_1)Z$.
\end{itemize}

Let $\pi \in \Funct(\mathcal P,\mathcal M)$ be the functor defined by
\begin{equation}
\forall Z \in \Obj(\mathcal P), \quad \pi(Z) = Z(Z^\dagger Z)^{-1} Z^\dagger \in G_m(\mathbb C^n)
\end{equation}
and
\begin{equation}
\forall (f,Z) \in \Morph(\mathcal P), \quad \pi(f,Z) =  \varpi(f)
\end{equation}
 $\mathcal P$ constitutes a principal 2-bundle over $\mathcal M$ with projection functor $\pi$. Its structure groupoid $\mathcal G$ is constituted by $G \simeq GL(m,\mathbb C)$ the group of matrices representing the basis changes on $\mathbb C^m$, and $H \simeq GL(m,\mathbb C)$ the group of matrices representing the rank $m$ linear operators of $\mathcal H$. $t$ is then the isomorphism between $H$ and $G$, and $\alpha$ is the conjugation. We can then defined the local trivialization equivalences of $\mathcal P$:
\begin{equation}
\forall P\in G_m(\mathbb C^n), g \in G, \quad \phi^i(P,g) = Z^i_0 g
\end{equation}
with $Z^i_0$ the coordinates matrix associated with $P$.
\begin{equation}
\forall Z \in \Obj(\mathcal P), \quad \bar \phi^i(Z) = (Z(Z^\dagger Z)^{-1}Z^\dagger,({Z^i_0}^\dagger Z^i_0)^{-1} {Z^i_0}^\dagger Z)
\end{equation}
with $Z^i_0$ the coordinates matrix of $Z(Z^\dagger Z)^{-1}Z^\dagger$. $({Z^i_0}^\dagger Z^i_0)^{-1} {Z^i_0}^\dagger Z$ is the passage matrix between the basis represented by $Z^i_0$ and the basis represented by $Z$.
\begin{eqnarray}
& & \forall \Omega \in \Morph(\mathcal M),(h,g)\in H \rtimes G, \nonumber\\
& &  \phi^i(\Omega,h,g) = (Z^i_{0q}h({Z^i_{0q-1}}^\dagger Z^i_{0q-1})^{-1} {Z^i_{0q-1}}^\dagger, Z^i_{0q-1} ({Z^i_{0q-2}}^\dagger Z^i_{0q-2})^{-1} {Z^i_{0q-2}}^\dagger,... \nonumber \\
& & \qquad ...,Z^i_{02} ({Z^i_{01}}^\dagger Z^i_{01})^{-1} {Z^i_{01}}^\dagger,Z_{01}^i g)
\end{eqnarray}
where $\Omega = P_q(P_{q-1}P_qP_{q-1})^{-1}...(P_2P_1P_2)^{-1}$ with $Z^i_{0j}$ the coordinates matrix of $\Ran P_j$.
\begin{eqnarray}
& & \forall (f,Z) \in \Morph(\mathcal P), \nonumber \\
& & \bar \phi^i(f,Z) = (\varpi(f), ({W_0^i}^\dagger W_0^i)^{-1} {W_0^i}^\dagger \chi(f) Z_0^i,({Z_0^i}^\dagger Z_0^i)^{-1} {Z_0^i}^\dagger Z)
\end{eqnarray}
where $Z_0^i$ is the coordinates matrix of $P_{\ker \chi(f)^\bot}$ and $W_0^i$ is the coordinates matrix of $P_{\Ran \chi(f)}$. We have well $t \bar \phi^i(f,Z) = \bar \phi^i(t(f,Z))$ since
\begin{equation}
({W^i_0}^\dagger W_0^i)^{-1} {W_0^i}^\dagger f \underbrace{Z_0^i({Z_0^i}^\dagger Z_0^i)^{-1} {Z_0^i}^\dagger}_P Z = ({W^i_0}^\dagger W_0^i)^{-1} {W_0^i}^\dagger f Z
\end{equation}
$\mathcal P$ is a categorical generalization of the Stiefel bundle, the classifying universal bundle for the $GL(m,\mathbb C)$-principal bundles (see \cite{rohlin}).\\
By definition the $G$-transition functions of $\mathcal P$ are such that
\begin{equation}
\forall P \in U^i \cap U^j, \quad \bar \phi^i \phi^j(P,e_G) = (P, ({Z^i_0}^\dagger Z^i_0)^{-1} {Z^i_0}^\dagger Z_0^j) = (P,g^{ij}(P))
\end{equation}
where $Z^i_0$ is the coordinates matrix of $P$. We have then 
\begin{equation}
g^{ij}(P) = ({Z^i_0}^\dagger Z^i_0)^{-1} {Z^i_0}^\dagger Z_0^j
\end{equation}
The $H$-transition functions are such that $\forall \Omega \in \Morph(\mathcal P)$
\begin{eqnarray}
\bar \phi^i \phi^j(\Omega,e_H,e_G) & = & (\Omega, ({W^i_0}^\dagger W^i_0)^{-1} {W^i_0}^\dagger W_0^j ({Z_0^j}^\dagger Z_0^j)^{-1} {Z_0^j}^\dagger Z_0^i, \nonumber \\
& & \qquad ({Z_0^i}^\dagger Z_0^i)^{-1}{Z_0^i}^\dagger Z_0^j) \\
& = & (\Omega, h^{ij}(Q,P), g^{ij}(P))
\end{eqnarray}
where $Z_0^i$ is the coordinates matrix of $P_{\ker \Omega^\bot}$ and $W_0^i$ is the coordinates matrix of $P_{\Ran \Omega}$. We have then
\begin{equation}
h^{ij}(Q,P) = ({W^i_0}^\dagger W^i_0)^{-1} {W^i_0}^\dagger W_0^j ({Z_0^j}^\dagger Z_0^j)^{-1} {Z_0^j}^\dagger Z_0^i
\end{equation}
The relation between the $H$-transition functions and the $G$-transition functions is well satisfied:
\begin{eqnarray}
h^{ij}(Q,P) g^{ij}(P) & = & g^{ij}(Q) ({Z_0^j}^\dagger Z_0^j)^{-1} {Z_0^j}^\dagger \underbrace{Z_0^i ({Z_0^i}^\dagger Z_0^i)^{-1}{Z_0^i}^\dagger}_{P} Z_0^j \\
& = & g^{ij}(Q) ({Z_0^j}^\dagger Z_0^j)^{-1} {Z_0^j}^\dagger Z_0^j \\
& = & g^{ij}(Q)
\end{eqnarray}
$\mathcal P$ is trivial in the sense where $h^{ijk}(P) = e_H$ since
\begin{eqnarray}
g^{ij}(P) g^{jk}(P) & = & ({Z_0^i}^\dagger Z_0^i)^{-1}{Z_0^i}^\dagger \underbrace{Z_0^j ({Z_0^j}^\dagger Z_0^j)^{-1}{Z_0^j}^\dagger}_{P} Z_0^k \\
& = & ({Z_0^i}^\dagger Z_0^i)^{-1}{Z_0^i}^\dagger Z_0^k \\
& = & g^{ik}(P)
\end{eqnarray}

\subsection{The 2-connection associated with the geometric phases}
The object-bundle of $\mathcal P$ is the Stiefel bundle, we can endow it with its natural connection (the universal connection of the $GL(m,\mathbb C)$-principal bundles, see the Narasimhan-Ramaman theorems \cite{narasimhan1,narasimhan2}), which defines the following $G$-gauge potential:
\begin{equation}
A^i(P) = ({Z_0^i}^\dagger Z_0^i)^{-1} {Z_0^i}^\dagger d Z_0^i \in \Omega^1(G_m(\mathcal H), \mathfrak g)
\end{equation}
Moreover we endow the arrow-bundle with a connection defining the following $H$-gauge potential:
\begin{equation}
\eta^i(Q,P) = ({Z_0^i}^\dagger Z_0^i)^{-1} {Z_0^i}^\dagger \left(\Omega^{-1} \dd \Omega\right) Z_0^i \in \Omega^1(G_m(\mathcal H)^2_{/ \mathcal R},\mathfrak h)
\end{equation}
with $\Omega = Q(PQP)^{-1}$ and $\Omega^{-1} = PQ$. Since $\Omega Z_0^i$ is a matrix representing a basis of $\Ran Q$, $\exists g \in G$ such that $W^i_0 g^{-1} = \Omega Z_0^i$. We have then
\begin{equation}
A^i(Q) = g^{-1} \eta^i(Q,P) g + g^{-1} A^i(P) g + g^{-1} dg
\end{equation}
The relation between the $G$-gauge potential and the $H$-gauge potential is then well satisfied (up to a $G$-gauge change).

\subsection{Horizontal lifts and parallel transport}
Let $t \mapsto P(t) \in G_m(\mathcal H)$ be a solution of the Schr\"odinger-von Neumann equation and $t \mapsto P_0(t) \in G_m(\mathcal H)$ be an eigenprojector of the Hamiltonian. We suppose that the almost adiabatic condition is satisfied, $\forall t$, $\dist(P(t),P_0(t))< \frac{\pi}{2}$. To simplify, we suppose also that $\forall t$, $P(t),Q(t) \in U^i$. The generalized time-dependent operator $t \mapsto \Omega(t) = P(t)(P_0(t)P(t)P_0(t))^{-1}$ constitutes an elementary pseudosurface of $\mathcal M$. The horizontal lift of $\Omega$ is then
\begin{eqnarray}
\mathcal H\ell^i(\Omega) & = & \PP_\Omega e^{- \int_{(P_0(0),P(0))}^{(P_0(t),P(t))} (A^i + \eta^i)} \\
& = & \Te^{- \int_0^t \tilde A^i(t')dt' - \int_0^t \tilde \eta^i(t')}
\end{eqnarray}
where 
$$\tilde A^i(t) = ({Z_0^i}(t)^\dagger Z_0^i(t))^{-1} {Z_0^i}(t)^\dagger \partial_t Z_0^i(t)$$
 and 
$$\tilde \eta^i(t) = ({Z_0^i}(t)^\dagger Z_0^i(t))^{-1} {Z_0^i}(t)^\dagger \Omega^{-1}(t) \dot \Omega(t) Z_0^i(t)$$
 $Z_0^i(t)$ being the coordinates matrix of $P_0(t)$. By applying the intermediate representation theorem (\cite{messiah}) on eqn. (\ref{wowf}) we have
\begin{eqnarray}
\psi(t) & = & \sum_{b=1}^m \left[ \Te^{- \ihbar^{-1} \hat E^{eff}(t')dt'} \Te^{- \int_0^t A(t')dt' - \int_0^t \eta(t')dt'} \right]_{ba} \Omega(t)\phi_{0b}(t) \\
& = & \sum_{b=1}^m \left[ \Te^{- \ihbar^{-1} \hat E^{eff}(t')dt'} g^i(t) \Te^{- \int_0^t \tilde A^i(t')dt' - \int_0^t \tilde \eta^i(t')dt'}g^i(t)^{-1} \right]_{ba} \nonumber \\
& & \qquad \times \Omega(t)\phi_{0b}(t) \\
& = & \sum_{b=1}^m \left[ \Te^{- \ihbar^{-1} \hat E^{eff}(t')dt'} g^i(t) \mathcal H\ell^i(\Omega) g^i(t)^{-1} \right]_{ba} \Omega(t)\phi_{0b}(t)
\end{eqnarray}
where $Z_0(t) = Z_0^i(t) g^i(t)$ ($Z_0(t)$ is the matrix representing the eigenvectors of $H^{eff}$). The geometric phases of an almost adiabatic quantum dynamics is then the horizontal lift of the pseudosurface defined by the generalized time-dependent wave operator. The formula (\ref{wowf}) can be then interpreted as being the parallel transport of $\phi_{0a}(0)$ along the pseudosurface $\Omega$ in the associated ``vector 2-bundle'' $\varpi : \mathcal E \to \mathcal M$, and modified by the conjugated dynamical phase $\Te^{- \ihbar^{-1} \hat E^{eff}(t')dt'}$:
$$ \hat E^{eff}(t) = \Te^{- \int_0^t (A(t') + \eta(t'))dt'} E^{eff}(t) \Te^{+ \int_0^t (A(t') + \eta(t'))dt'} $$
Finally we can note that the use of an usual time-dependent wave operator (with $\dot P_0 = 0$) is just a particular case of the present discussion with a pinched pseudosurface.

\section{Conclusion}
The categorical bundle structure is extended to the case where the base space is not a trivial category but an affine 2-space. The new strucure permits to define the horizontal lifts of objects called the pseudosurfaces. For an impervious pseudosurface, we recover the horizontal lifts of the usual surfaces (the surface of the firts kind supported by the pseudosurface) previously studied by different authors \cite{laurent,breen,kalkkinen2,aschieri2,viennot5,baez1,baez2,baez3,wockel}, but the notion of pseudosurface is more general. The condition $t^{Lie}(\eta^i(y,x)) = A^i(y)-A^i(x)$ implies that $A^i \in \Omega^1(U^i,t^{Lie}(\mathfrak h))$ (if $U^i$ is totally linkable) and then that $a^i$ the connection of the quotient bundle $R$ must be pure gauge. This reduces the possible applications of the present work. The most interesting cases are in these conditions, like the example presented section 6, such that $t(H)=G$ (i.e. $t$ is a surjective homomorphism) and especially when $H=G$ and when $t$ is an automorphism of $G$. Another example of this kind can be found in \cite{viennot8} where the space of the density matrices endowed with a group action plays the role of an Euclidean affine 2-space. Moreover we can have a non-trivial connection $a^i$ on the quotient bundle $R$ if a part of the objects of the base 2-space are linkable only to themselves. The fact that the wave operators of the quantum dynamics can be viewed as (not impervious) pseudosurfaces augurs the futur developpement of new kinds of non-abelian geometric phases for quantum systems, particularly for the non-hermitian quantum systems where the wave operator seems play an important role \cite{viennot2}. Moreover the possibility to study the new physical theories (as the string and brane theory) in the framework of this generalization could be interesting since some attempts to develop a categorical theory of quantum gravity have been proposed \cite{isham,raptis}.

\end{document}